\documentclass[aps, prb, onecolumn, nofootinbib, citeautoscript, 10pt]{revtex4-2}

\usepackage[papersize={8.5in,11in}]{geometry}

\usepackage[colorlinks=true]{hyperref}
\hypersetup{
	bookmarks=true,         
	unicode=false,          
	pdftoolbar=true,        
	pdfmenubar=true,        
	pdffitwindow=false,     
	pdfstartview={FitH},    
	pdfkeywords={keyword1} {key2} {key3}, 
	pdfnewwindow=true,      
	colorlinks=true,       
	linkcolor=blue,           
	citecolor=blue,        
	filecolor=magenta,      
	urlcolor=blue           
} 

\usepackage{amsmath,amssymb, ulem, float} 
\usepackage{comment, cancel}
\usepackage[version=4]{mhchem}
\usepackage{dcolumn}
\usepackage[tight]{subfigure} 
\usepackage[dvipsnames]{xcolor} 
\usepackage{color, enumitem}
\usepackage{amssymb,amsmath}
\usepackage{tabularx, graphicx}
\usepackage{bbm,bm,array,physics}
\usepackage{dsfont}

\def \nn{\nonumber \\}
\def\*#1{\mathbf{#1}} 

\newcommand{\bk}{\mathbf{k}}
\newcommand{\bp}{\mathbf{p}}
\newcommand{\bq}{\mathbf{q}}
\newcommand{\br}{\mathbf{r}}
\newcommand{\bOmega}{\bar{\Omega}}
\newcommand{\bbq}{|\bar{\mathbf{q}}|}

\begin{document} 
	
\title{Generic deformation channels for critical Fermi surfaces including the impact of collisions}

\author{Kazi Ranjibul Islam}
\affiliation{School of Physics and Astronomy and William I. Fine Theoretical Physics Institute,
	University of Minnesota, Minneapolis, MN 55455, USA}

\author{Aditya Savanur}
\affiliation{School of Physics and Astronomy and William I. Fine Theoretical Physics Institute,
	University of Minnesota, Minneapolis, MN 55455, USA}

\author{Ipsita Mandal}
\email{ipsita.mandal@snu.edu.in}
\affiliation{Department of Physics, Shiv Nadar Institution of Eminence (SNIoE), Gautam Buddha Nagar, Uttar Pradesh 201314, India}

\begin{abstract}
This paper constitutes a sequel to our theoretical efforts to determine the nature of generic low-energy deformations of the Fermi surface of a quantum-critical metal, which arises at the stable non-Fermi liquid (NFL) fixed point of a quantum phase transition. The emergent \textit{critical Fermi surface}, arising right at the Ising-nematic quantum critical point (QCP), is a paradigmatic example where an NFL behaviour is induced by the strong interactions of the fermionic degrees of freedom with those of the bosonic order parameter. It is an artifact of the bosonic modes becoming massless at the QCP, thus undergoing Landau-damping at the level of one-loop self-energy. We resort to the well-tested formalism of the quantum Boltzmann equations (QBEs) for identifying the excitations. While in our earlier works, we have focused on the collisionless regime by neglecting the collision integral and assuming the bosons to be in equilibrium, here we embark on a full analysis. In particular, we take into account the bosonic part of the QBEs as well, which, however, turn out to have no effect on the solutions. Decomposing the master equation into angular-momentum ($\ell$) channels, the emergent modes are of two types: Fermi-surface deformations with discrete spectra and particle-hole excitations forming a continuous band. The long-lived zero-sound mode, which corresponds to $\ell = 0$, is found to be robust against damping effects. Intriguingly, we have an infinite family of discrete modes corresponding to higher-order harmonics of the net deformation.
\end{abstract}

\maketitle

\tableofcontents

\section{Introduction}

Non-Fermi liquid (NFL) metals are one of the most prominent representatives of the consequences of strong correlations in condensed matter systems \cite{metlsach1, metlsach, olav, Lee-Dalid, ips-uv-ir1, ips-uv-ir2, ips-fflo, ips-nfl-u1,  metzner, ips-2kf, ips-cavity, ips-rafael, ips-cdw-moire}, defying analytical investigations using conventional theories. In particular, Landau's Fermi-liquid theory, which explains normal metals extremely well, breaks down for the case of NFLs, caused by the destruction of the Landau quasiparticles. Consequently, they possess \textit{strange} thermodynamical and transport properties \cite{ips-subir, ips-c2, ips-hermann, ips-hermann2, ips-hermann3, ips-hermann-review}, being starkly distinct from those of the normal metals. One such characteristic involves the resistivity not having a quadratic-in-$T$ dependence, which is a hallmark property of Fermi liquids (describing normal metals). Other such distinctions can be detected in optical conductivity \cite{ips-subir, ips-c2} and increased affinity towards superconducting instability \cite{ips2, ips-sc, ips-c2}. Some well-known scenarios, where such \textit{strange-metallic} characteristics emerge, happen to be (1) finite-density fermions interacting with order-parameter bosons that become massless at a quantum critical point (QCP) \cite{metlsach1, metlsach, Lee-Dalid,ips-uv-ir1, ips-uv-ir2, ips-fflo, ips-rafael, ips-2kf, ips-cavity}; (2) finite-density fermions interacting with transverse gauge field(s) \cite{olav, ips2, ips3, ips-nfl-u1}; (3) chemical potential tuned to the band-touching points of semimetals in the presence of unscreened Coulomb interactions \cite{Abrikosov, moon-xu, rahul-sid, ips-rahul, malcolm-bitan, ips-birefringent}. In this paper, our focus will be on the stable NFL arising at the two-dimensional (2d) Ising-nematic QCP \cite{metlsach1, metlsach, Lee-Dalid, ips-uv-ir1, ips-uv-ir2}.

There have been extensive efforts to obtain perturbative NFL states implementing distinct and complementary ways, for which we provide an overview. The first popular attempt was the introduction of a large number ($N$) of flavours \cite{ALTSHULER,polchinski,ybkim} as a mathematical tool, employing the natural intuition that extra flavours do not introduce qualitatively new element to the theory, except that the flavour-symmetry group is enlarged. However, eventually, it was established beyond question that the $N \rightarrow \infty$ limit for finite-density fermions is not described by a mean-field theory \cite{SSLee, metlsach1}. The physical reason behind this is the fact that there exists a large amount of residual quantum-fluctuations of the Fermi surface. Following these seminal works, controlled approximations were formulated which were not based on expansion in $1/N$.
One methodology was deforming the dynamics of the theory, for example by modifying the dispersion of a critical boson, to suppress quantum fluctuations at low-energies \cite{nayak, mross}. In another effort, accessing two-dimensional NFLs was made possible by increasing the number of one-dimensional chains, where bosonization provides a controlled analytical tool \cite{Jiang}.
Finally, the mathematical control of perturbation was gained by an intermediate step of modifying the number of dimensions of spacetime (in which the system is embedded) in a continuous manner. This process can be implemented by extending either Fermi surface's dimension \cite{olav} or its co-dimension \cite{senshank}. In this paper, we adopt the framework where a controlled approximation was devised via a dimensional regularization scheme, where the co-dimension of Fermi surface is extended to access the NFLs \cite{Lee-Dalid, ips-uv-ir1}.

Ising-nematic ordering implies the spontaneous breaking of the rotational symmetry of a Fermi surface in the $xy$-plane, which describes the Pomeranchuk instability in the angular-momentum channel with the value $\ell=2$. Consequently, a fourfold symmetric Fermi surface transitions into a twofold symmetric one, causing the $x$- and $y$-directions to become anisotropic \cite{metlsach1, sachdev_2011}. This broken symmetry is explained by invoking the Ising-nematic order parameter, which has the same Ginzburg-Landau action as the Ising order parameter of magnetism. Hence, this is captured by a real scalar bosonic field ($\phi$), whose momentum is centred at the wavevector $\mathbf{Q} = 0$. For the Ising-nematic QCP, a sharp Fermi surface is retained at the quantum phase transition, although the Landau quasiparticles get destroyed \cite{Lee-Dalid,ips-uv-ir1, ips-uv-ir2, ips-subir}. Because the Ising-nematic phase is predicted to exist in various high-T$_c$ materials like cuprate superconductors \cite{yoichi, hinkov, Kohsaka, Daou} and Fe-based superconductors \cite{hsu2008superconductivity, margadonna2008crystal,mcqueen2009tetragonal}, it can be viewed as a simple model system to study generic features of NFLs arising at QCPs \cite{Lee-Dalid, ips-uv-ir1, ips-fflo, ips-2kf, ips-cavity}. Since the Fermi surface continues to persist, the pertinent question that we ask is what the dispersion relations of the collective modes/excitations of the Fermi surface's deformations look like, when we decompose the displacements into angular momentum channels. In our earlier works, we attempted to answer this question by assuming the collisionless limit \cite{ips-zero-mode, ips-kazi}. Various other works have also investigated these aspects using various mathematical arguments and numerical simulations \cite{debanjan, else, khveshchenko}.

In order to deal with the challenging case of NFLs, we will use the nonequilibrium Green's function technique, outlined in Refs.~\cite{prange, kim_qbe}, and used in our earlier analyses \cite{ips-zero-mode, ips-kazi}. This allows us to derive the quantum Boltzmann equations (QBEs), giving rise to the \textit{so-called} generalised Landau-interaction functions (GLFs). The GLFs appear in the linearised kinetic equations as the analogues of the Landau parameters (defined for the Fermi-liquid case). Unlike the Landau parameters though, the GLFs have a frequency-dependence (in addition to the usual angular-dependence). The frequency-dependence is caused by the Landau-damping that the gapless bosonic modes undergo. In our preceding works, we have focused on the collisionless regime (by neglecting the collision integral) and, also, assumed the bosons to be in equilibrium. Here, we improve upon that methodology by embarking on a full-fledged analysis. In particular, we take into account the bosonic part of the QBEs and also include all the collision terms arising for each kind of particle. Overall, we would like to emphasise that the primary new advancement of our work is twofold: (1) the systematic inclusion of the \textit{collision integral part} within the QBE framework, going beyond the collisionless analysis of earlier works (viz. Refs.~\cite{ips-zero-mode, ips-kazi}); and (2) studying the frequency-dependent and complex GLFs, encoding damping effects to give a full-fledged description of generic deformation channels for critical Fermi surfaces.

The paper is organized as follows: In Sec.~\ref{secmodel}, we introduce the model for the Ising-nematic quantum critical point, which shows an NFL fixed point \cite{Lee-Dalid, ips-uv-ir1, ips-uv-ir2, ips-subir,ips-sc, ips-zero-mode, ips-kazi}. We also review the form of the Green's functions in the Keldysh formalism, which is useful for setting up the QBEs for the fermions and the bosons. In Sec.~\ref{secqbe}, we demonstrate the forms of the QBEs, including the relevant collision integrals, considering two scenarios: (A) the bosons are assumed to be in equilibrium, with no bosonic kinetic equation as a consequence; (B) the bosons are slightly away from equilibrium, giving rise to coupled QBEs governing the dynamics of both the fermions and the bosons. Sec.~\ref{secsol} deals with obtaining the explicit solutions by decomposing our equations into distinct angular-momentum channels. The $\ell = 0$ mode represents the zero sound, and can be easily understood analytically, as explained in Sec.~\ref{seczero}. However, we also obtain the solutions retaining generic $ \ell$-values, which have to be implemented numerically, due to the complexity of the coupled equations involved. This we take care of in Sec.~\ref{seccode}. We conclude with a summary and outlook in Sec.~\ref{secsum}. The appendices are devoted to explaining some notations and generic ingredients of the Keldysh framework.

\section{Model}
\label{secmodel}

For describing the Ising-nematic QCP, there are strong interactions between two kinds of degrees of freedom --- the itinerant fermionic excitations (around the critical Fermi surface) and the modes of the critical boson (which are essentially massless right at the phase transition). A stable NFL phase arises right at the QCP as a consequence of their mutual interactions, decimating the Landau-quasiparticle nature of the fermions. The effective field theory, which is able to capture all this behaviour, is described below \cite{metlsach1,sachdev_2011}:
\begin{enumerate}[leftmargin=*]

\item First, the Ginzburg-Landau action for order parameter in imaginary time ($\tau$), which represents a real scalar bosonic field ($ \phi $), is given by
\begin{align}
	\label{ac1}
	S_b & =  \frac{1}{2} \int d\tau \,dx \, dy\left[  \phi(\tau,x,y)
	\left(- \partial_\tau ^2 -  c_\phi ^2 \, \partial_x ^2 -  c_\phi ^2 \, \partial_y ^2
	\right) \phi (\tau,x,y)
	+  r_\phi \,\phi^2 (\tau,x,y)
	+  \frac{u_\phi \,\phi^ 4(\tau,x,y) } {12}
	\right].
\end{align}
Here, the position space is spanned by the $(x,\,y)$ coordinates, $c_\phi$ is the boson velocity, $ r_\phi$ is the parameter tuning across the QCP, and $u_\phi$ is the coupling constant for the $\phi^4$-term. Computing the tree-level/engineering dimensions of the various terms, all couplings can be scaled away or set equal to unity, except for $r_\phi$. 
The QCP is the zero-temperature phase transition appearing at $ r_\phi =0$. The physics of this purely bosonic part, however, is not the complete story, as the fermions coupling to the gapless bosons at $ r_\phi =0$ change the nature of the quantum critical fluctuations. The resulting composite system is strongly coupled, with the bosons acquiring Landau-damping, causing the fermions to behave as an NFL in turn. Although the NFL fixed point emerges at temperature $T=0$, which of course is not observable, the quantum effects show their distinctive NFL-like features  in an extended fan-shaped quantum critical region emanating from the QCP \cite{sachdev_2011}.

\item For the purely fermionic part, one of the well-tested approaches is to incorporate the patch theory, introduced in Refs.~\cite{metlsach1,sachdev_2011,Lee-Dalid,ips-uv-ir1,ips-uv-ir2,ips-nfl-u1}. The usefulness of using these coordinates lies in the fact that the itinerant fermions, in the vicinity of a given point on the Fermi surface, primarily couple with the bosons whose momentum ($\mathbf q$) is in a direction perpendicular to the local Fermi momentum. Consequently, the fermions in different patches on the Fermi surface are effectively decoupled from each other in the low-energy limit, except of course when they have antiparallel tangents. It makes it possible to extract all the physical properties (e.g., Green’s functions) from the information contained in the local patches, without having to refer to the global properties of the Fermi surface. Using the patch coordinates, the kinetic action for the fermionic excitations is captured by
\begin{align}
	\label{ac2}
	S_f  & =   
	\int_{k} \psi^\dagger (k) 
	\left ( -i \,k_0  +  \delta_k \right ) \psi (k) \,,
	\quad \delta_k =  v_F \,k_{\perp}  +   \frac{k_{\parallel}^2} {2\,m}\,,
\end{align}
in the Fourier space. Here, $\psi$ denotes the fermionic field residing on the patch under consideration. We have used the condensed notations of $k \equiv (k_0, \mathbf k)$ (with $k_0 $ denoting the Matsubara frequency) and $ \int_k  \equiv \int \frac{dk_0\,dk_{\parallel}\,dk_\perp } {(2\,\pi)^3}$.
While writing the kinetic terms in the the patch coordinates, we have expanded the fermion momentum about the local Fermi momentum, $k_F$, such that $ k_{\perp}$ is directed perpendicular to the local Fermi surface and $k_{\parallel}$ is tangential to it.

\item The last bit comprises the coupling terms between the bosons and the fermions, which are captured by
\begin{align}
	\label{ac3}
	S_{\rm int} = \tilde e
	\int_{k} \int_q \left(  \cos k_x -\cos k_y  \right)
	\phi(q) \,\psi^\dagger (k + q) \, \psi(k) \,,
\end{align}
where $\tilde e$ is the fermion-boson coupling constant. We would like to point out that this piece of the total action is written in the global momentum space coordinates (rather than the patch coordinates of the Fermi surface).
Converting the above to the patch coordinates, and keeping only the leading order terms in momentum about the Fermi surface, we get
$\left(  \cos k_x -\cos k_y  \right) \simeq \cos k_F$. Re-expressing the coupling strength as $e = \tilde e \, \cos k_F$, we then get the following form for the fermion-boson interaction in the patch coordinates:
\begin{align}
\tilde S_{\rm int} 
= e\,  \int_{k}  \int_q \phi(q) \,\psi^\dagger (k + q) \, \psi(k) \,.
\end{align}

\end{enumerate} 

The penultimate step for obtaining the final form of the total action, $S_{\rm tot}$, involves an appropriate rescaling of the energy and momenta, and dropping the irrelevant terms \cite{Lee-Dalid,ips-uv-ir1} after determining the engineering dimension of each term. This is implemented by setting $[k_0] = [k_\perp] = 1$ and $ [k_\parallel] =1/2$. This finally leads to
\begin{align}
S_{\rm tot} & = S_f  +  \tilde S_b  +  \tilde S_{\rm int} \,,\quad
\tilde S_b  = \frac{1}{2} \int_{k}\,k_{\parallel}^2 \,\phi(-k) \,\phi(k) \,.
\end{align}
The patch coordinates enables us to extract the correct forms of the one-loop self-energies for the bosons and the fermions \cite{metlsach1, Lee-Dalid, ips-uv-ir1}, as described below:
\begin{enumerate}[leftmargin=*]

\item
From $S_{\rm tot}$, the bare Matsubara Green's functions for the fermions and bosons are found to be
\begin{align}
	\label{eqbareGF}
	G^{(0)} (k)
	= \frac{1} {  i \, \, k_0 - \delta_k  } \text{ and }
	D^{(0)} (q) = \frac{1} { - \, q_{||}^2} \,,
\end{align}
respectively. 
	
\item Since the bare kinetic term of the boson depends only on $q_{||}$, we
need to include the lowest-order quantum corrections to ensure that the loop-integrals are infrared- and ultraviolet-finite \cite{Lee-Dalid,ips-uv-ir1}. Since the one-loop bosonic self-energy evaluates to
$ \Pi(q)= - e^2 \int_k G^{(0)}(k + q)
\,G^{(0)} (k)=\frac{e^2\, m \,|q_0|}
{2 \, \pi \, v_F \,|q_{||}|}$,
we use the one-loop corrected bosonic propagator,
\begin{align}
	D^{(1)}(q)=\left[
	\left(  D^{(0)} (q)\right) ^{-1} 
	-\Pi(q)   \right]^{-1}
	= - \,\frac{1} {q_{||}^2  +  \frac{e^2  \, m\, |q_0|} {2 \,\pi \,v_F \,|q_{||}|}} \,,
	\label{eqbareD}
\end{align} 
in all our subsequent loop calculations.
We note that $ \Pi(q)$ takes a form which represents an overdamped harmonic oscillator, thus earning it the nomenclature of \textit{Landau-damping}. The validity of this expression corresponds to the limits of $| q_0|/|q_\parallel| \ll 1 $, $|\mathbf q| \ll k_F$, and $ |\mathbf q| \rightarrow 0 $.

\item Using the dressed bosonic propagator (containing the all-important effects of Landau-damping), the one-loop fermion self-energy in the Matsubara space turns out to be 
\begin{align}
	\Sigma(k)= -\, \frac{   i \, \,   e^{\frac{4} {3}}\, \text{sgn}(k_0) \, |k_0|^{ \frac{2} {3} }}
	{2 \, \sqrt{3}\, \pi^{ \frac{2} {3} }\,k_F^{\frac{1} {2}}} \,.
	\label{selfenergyform}
\end{align}
The imaginary part of the self-energy represents the lifetime of the underlying quasiparticles. For a Fermi liquid, the self-energy turns out to be proportional to $\omega^2$, which is much smaller than the excitation energy $\omega$, in the limit $\omega \rightarrow 0 $. This is the reason why we can define long-lived Landau quasiparticles for a normal metal (or Fermi liquid). However, here we find that $\text{Im}\Sigma \propto |\omega|^{ \frac{2} {3} }$, whose magnitude is much greater than $|\omega|$ in the limit $\omega \rightarrow 0$. Thus the quasiparticle picture fails for the Ising-nematic QCP, making the traditional Boltzmann formalism inadequate for deriving the corresponding kinetic equations. 
	
\end{enumerate}

\subsection{Keldysh formalism}
\label{seckeldysh}

To deal with a generic nonequilibrium situation, a widely used method is to use the closed-time Keldysh contour \cite{kamenev}. Using the field operators on the forward and backward branches of the time contour, we write down the total Keldysh action, as shown in Refs.~\cite{ips-zero-mode, ips-kazi}, which we do not repeat here. 
The Keldysh formalism comes to include various types of Green’s functions, which appear as a consequence of the respective locations of their time arguments on the folded contour. These are shown in Appendex~\ref{app1} for the sake of completeness. 
Using these ingredients, we finally change the spacetime coordinates to $(t_{\rm rel}=t_1-t_2, \, \mathbf r_{\rm rel} =\mathbf r_1 -\mathbf r_2)$ and $(t=\frac{t_1 + t_2} {2}, \,  \mathbf r  = \frac{ \mathbf r_1 +  \mathbf r_2} {2})$, which are also known as the Wigner coordinates. While the former refers to the \textit{relative coordinates}, the latter comprises the \textit{centre-of-mass} description. Through this step, we can easily describe the system in the equilibrium limit, because the dependence on $(t, \mathbf r)$ drops out due to the emergent translational invariance both in time and space, causing all the Green's functions to depend \textit{only} on $(t_{\rm rel}, \mathbf r_{\rm rel})$. Defining $k_1\equiv (\omega_{k_1}, \, \mathbf k_1)$ and $k_2\equiv (\omega_{k_2}, \, \mathbf k_2)$ to be the canonically conjugate energy-momentum variables of $x_1\equiv(t_1,\mathbf r_1)$ and $x_2\equiv(t_2,\mathbf r_2)$, respectively, we can then easily perform our calculations by going to the Fourier space, with the canonically conjugate variable defined as $ k =\frac{ k_1 - k_2}{2} $.
In a generic nonequilibrium situation, the additional dependence on $(t_{\rm rel}, \, \mathbf r_{\rm rel})$ needs to be accounted for, when we can use the canonically conjugate variable, $  q=  k_1  +   k_2$.
For the sake of notational simplification, we define the shorthand symbols of $ x_{\rm rel}  = 
(t_{\rm rel}, \, \mathbf r_{\rm rel})$, $ k  \equiv (\omega_k, \, {\mathbf k})$ and $
q \equiv (  \omega_q, \,\mathbf q)$ [and set $\omega_q=\Omega $], combining everything into three-vectors.

\subsection{Equilibrium properties and generalised distribution functions} 

For our system, the explicit expressions for the bare (aka noninteracting) Green's functions, at equilibrium, are given by
\begin{align}
	\label{listgf}
	G_{bare}^{R}( \omega_k ,\mathbf{k})  & =  
	G^{(0)}( i \, \,k_0,\mathbf k) \Big \vert_  { i \, \,k_0 \rightarrow \omega_k  +  i \, \,0^ + }
	= \frac{1} {\omega_k- \delta_k  +  i \,0^ +  }\,,\nn
	G_{bare}^{A }(\omega_k ,\mathbf{k}) 
	& = G^{(0)}( i \, \,k_0,\mathbf k) \Big \vert_  { i \, \,k_0 \rightarrow \omega_k -  i \, \,0^ + } 
	=  \frac{1} {\omega_k- \delta_k - i \,0^ +  } \,,
\end{align}
and
\begin{align}
\label{listgfbos}
& D_{1}^{R}(\omega_k,\mathbf{k}) 
= D^{(1)}( i \, k_0,\mathbf k) \Big \vert_  { i \, k_0 \rightarrow \omega_k  +  i \, 0^ + }
= \frac{ 1}
{ - \, k_{\parallel}^2 - \Pi_0^R(\omega_k, \bk ) }
\,,\quad
D_{1}^{A}(\omega_k,\mathbf{k}) 
= D^{(1)}( i \, k_0,\mathbf k) \Big \vert_  { i \, k_0 \rightarrow \omega_k -  i \, \,0^ + }
= \frac{1}
{ - \, k_{\parallel}^2 + \Pi_0^R(\omega_k, \bk) }\,,
\nn & \Pi_0^R(\omega_k, \bk ) = -\, i \,\chi \, \frac{\omega} {|\bk|}\,, \quad 
\chi = \frac{ e^2\, m} {\pi \,  v_F } =  8 \,m \, k_F \,\alpha \,,\quad
\alpha = \frac{ e^2} {8 \,\pi  \, v_F\,k_F}\,,
\end{align}
for the fermions and the bosons, respectively. The superscripts ``$R$'' and ``$A$'' refer to the ``retarded'' and ''advanced'' ones, respectively. The expressions follow straightforwardly from Eqs.~\eqref{eqbareGF} and \eqref{eqbareD}.
While performing the analytic continuation, $ i \, k_0 \rightarrow \omega_k  +  i \, \,0^ + $,
one has to use
$$\text{sgn} [k_0] \equiv \text{sgn}\Big [\text{Im} [ i \, k_0] \Big ]
\rightarrow \text{sgn}\Big [\text{Im} \big [\omega_k +  i \,0^ +  \big ] \Big ]
=\text{sgn} \big [ 0^ +  \big ] =1\,, $$ and
an analogous relation for the case of $ i \, k_0 \rightarrow \omega_k -  i \, \,0^ + $, required to derive $D_{1}^{A}$.
We note that $\alpha $ represents the analogue of the fine-structure constant of the electromagnetic interactions.

The feedback of the overdamped bosons on the fermions is computed via the retarded and advanced fermion self-energies at one-loop order. The equilibrium expressions for these self-energies at one-loop order, after analytic continuation of Eq.~\eqref{selfenergyform} to real frequencies, are obtained as
\begin{align}
\label{eq-ferm-self-en}
\Sigma^{R}(\omega_k)
= - \frac{e^{4/3} \left[ \sqrt 3\,\text{sgn}[\omega_k]  +  i \right]
	|\omega_k |^{ \frac{2} {3} } } 
{ 4\, v_F\,\pi ^{ \frac{2} {3} } \left( m/v_F\right)^{ \frac{1} {3} } } \text{  and  }
\Sigma^{A}(\omega_k)
= - \frac{e^{4/3} \left[ \sqrt 3\,\text{sgn}[\omega_k] - i \right]
	|\omega_k |^{ \frac{2} {3} } } 
{ 4\, v_F\,\pi ^{ \frac{2} {3} } \left( m/v_F\right)^{ \frac{1} {3} } }\,,
\end{align}
respectively.
Thus, the one-loop-corrected Green's functions take the forms of
\begin{align}
	\left[ G_1^{R/A}(\omega_k, \mathbf{k}) \right]^{-1}
	= \left[ G_{bare}^{R/A}( \omega_k ,\mathbf{k})  \right]^{-1} -\Sigma^{R/A}(\omega_k)\,.
\end{align}
The equilibrium Green's functions are related to the spectral function ($A$) and the Fermi-Dirac distribution, $ f_0(\omega) = [ 1  +  \, e^{\beta \, \omega} ]^{-1} $ (at a temperature $T =1/\beta$), as [cf. Eq.~\eqref{eqspec} of Appendix~\ref{appeq}]
\begin{align}
	G^<(\omega_k,\mathbf k) & = i \, f_0(\omega_k) \, A( \omega_k,\mathbf k) \,,\quad
	G^>(\omega_k,\mathbf k) =-i \left [1- f_0(\omega_k) \right ] A(\omega_k,\mathbf k)\,,
\end{align}
where
\begin{align}
	A( \omega_k,\mathbf k) = -  2\, \text{Im}[G^R(\omega_k,\mathbf{k})]
	=\frac{ 2\, \text{Im} [\Sigma^R( \omega_k,\mathbf k)]}
	{ \Big [ \,\omega_k- \delta_k-\text{Re} [\Sigma^R ( \omega_k,\mathbf k)] \,\Big]^2
		+  \Big [ \, \text{Im} [\Sigma^R ( \omega_k,\mathbf k)] \,\Big]^2} \,.
\end{align}
Here, a few points are worth reiterating.
In Landau's Fermi-liquid theory, the imaginary part of the fermionic self-energy turns out to be $\text{Im}[\Sigma^R] \sim \omega_k^2 \ll |\omega_k| $ for $|\omega_k| \rightarrow 0 $.
As a result, the equilibrium spectral function $A$ takes the form of a sharply-peaked function of $\omega_k$, such that $ A(\omega_k,\mathbf{k}) 
\simeq 2 \,\pi \, \, \delta \big( \omega_k - \xi_\mathbf{k} - \text{Re}[\Sigma^R(\omega_k,\mathbf{k})] \big),$
where $ \xi_\mathbf{k}$ is the bare fermion dispersion.
This relation indicates that for fluctuations close to the equilibrium, we can construct a closed set of  equations for the fermion distribution function $ f(\omega_k,\mathbf k;t,\mathbf r)$, which constitute the QBEs. The set of linearised QBEs for the fluctuation
$ \delta f(\omega_k,  \mathbf k;t,\mathbf r) = f(\omega_k, \mathbf k;t,\mathbf r) - f_0(\omega_k)$ describes the transport equations for a Fermi liquid. On the contrary, for our NFL system, Eq.~\eqref{eq-ferm-self-en} gives $ \text{Im} [\Sigma^R] \propto |\omega_k |^{ \frac{2} {3} }$, implying that $A(\omega_k, \mathbf{k})$ is not a sharply-peaked function of $\omega_k $ at equilibrium, contrary to the behaviour of Fermi liquid systems. As a result, $  \delta f(\omega_k, \mathbf k;t,\mathbf r)$ does not satisfy a closed set of equations even at equilibrium. We thus need to devise a formalism which does not depend on the smallness of the decay rate, which is a quantity that is proportional to the width of the peak in $A (\omega_k, \mathbf{k}) $ as a function of $\omega_k$. Observing that $A (\omega_k, \mathbf{k})$ has a well-defined peak around $ \delta_k = 0$ \cite{prange,kim_qbe} (since $\Sigma^R$ is a function of $\omega_k $ only), and $\int_{-\infty}^{\infty} \frac{d\delta_k}{ 2\, \pi  } A (\omega_k, \mathbf{k}) = 1$, we conclude that $G^<$ and $G^>$ are sharply-peaked functions of $\delta_k$. Integrating over the region of the sharp peak, it is useful to define the generalised fermion distribution function $f$ (also known as a Wigner distribution function) as~\cite{kim_qbe, prange}
\begin{align}
	\label{eq:gdf}
	\int \frac{d\delta_k}{ 2\, \pi  } \,
	G^<(\omega_k, {\mathbf k} ; \omega_q,\mathbf q) 
	=  i \, \,f(\omega_k, {\mathbf k} ;\omega_q, \mathbf q)\,,\quad
	\int \frac{d\delta_k }{ 2\, \pi  } \,
	G^>(\omega_k, {\mathbf k} ; \omega_q,\mathbf q) 
	= 
	i   \left[\, f(\omega_k,{\mathbf k};\omega_q, \mathbf q) -1 \, \right] ,
\end{align}
which works in the absence of well-defined Landau quasiparticles. These relations will allow us to derive the set of QBEs which can characterise the fluctuations of a critical Fermi surface, as long as the system is not far away from the equilibrium.

Applying Eq.~\eqref{eqspec} to Eq.~\eqref{listgfbos}, we get the bosonic spectral function as
\begin{align}
B(\omega_k, \bk ) = 
\frac{2 \,\chi \, \omega_k \, |\bk|} {|\bk|^6  +  \chi^2 \, \omega_k^2} \,,
\label{eqbosspec}
\end{align}
which is peaked at the frequency values of $\omega= \pm \,\omega_0 (\bk) $ with $ \omega_0 (\bk) = |\bk|^3 / \chi$. As a result, unlike the fermionic sector, here we can integrate over $\omega_k $ just as one does in the standard Fermi liquid case (utilizing the fact that the quasiparticle distributions are sharply peaked). We approximate the bosonic spectral function as $B(\omega_k, \bk ) = \pi \left[\delta(\omega_k-\omega_0(\bk))-\delta(\omega_k  +  \omega_0(\bk)\right]
Z_B(\bk)$, where 
\begin{align}
	Z_B(\bk) \equiv   
	\int_0^{|\bk|} \frac{d\omega_k}{\pi} \, B(\omega_k, \bk ) = \frac{|\bk|}{\pi \,\chi}
	\ln \left(1  +  \frac{\chi^2}{|\bk|^2}\right)  .
	\label{eq_ZB_int}
\end{align}
We set the upper cut-off in Eq.~\eqref{eq_ZB_int} to $|\bk|$ assuming $\omega_0(\bk) << |\bk|$, and will show later that it does not influence the results. For a small deviation from the equilibrium situation, we can then define
\begin{align}
	\int_0^\infty \frac{d\omega_k} {\pi} D^<(\omega_k , \bk ;t, \br ) = 
	n \big ( \omega_0(\bk), \bk;t, \br \big ) \,Z_B(\bk) \text{ and }
	\int_0^\infty \frac{d\omega}{\pi} D^>(\omega_k, \bk ;t, \br ) = 
	\left [ 1 + n  \big (\omega_0(\bk),\bk;t, \br \big ) \right ] Z_B(\bk) \,,
	\label{eq-bos-dist}
\end{align}
where $  n \big (\omega_0(\bk), \bk ;t, \br \big ) $ is the number-density of the critical bosons, for the modes carrying momentum $\bk$ and frequency $\omega_0(\bk) $ at $(t, \br)$.

\section{Solutions to the quantum Boltzmann equations}
\label{secqbe}

In order to derive the QBEs for the collective modes of a critical Fermi surface, we need to refer to its global properties. For the sake of simplicity, we assume a circular Fermi surface with the Fermi momentum vector given by  ${\mathbf k}_F = k_F \, \hat{{\mathbf k}}_{\text{rad}}$, where $ \hat{{\mathbf k}}_{\text{rad}}$ is the angle-dependent unit vector pointing radially outward on the Fermi surface. Moreover, our derivations will be limited to the zero temperature limit (i.e., $T=0$).
We use the parametrization $\mathbf {k} =  \frac{\mathbf k_1 -\mathbf k_2} {2} 
\equiv \left( k_F + k_\perp \right) \hat{{\mathbf k}}_{\text{rad}} 
+ k_\parallel \,{\hat{ \mathbf {\theta}}}_{\mathbf k} $, $\mathbf q = \mathbf k_1 + \mathbf k_2$,  and $\theta_{\mathbf k} \,
( \theta_{\mathbf q} )$ is the angle that the vector $\mathbf {k} \,(\mathbf q)$ makes with the $x$-axis. Since we are focusing on small perturbations of a patch on the Fermi surface, with the local Fermi momentum $k_F$, we must have $ \left \lbrace |k_\perp|, |k_\parallel|,
|\mathbf q |  \right \rbrace \ll k_F$. The functional dependence of $f(\omega_k, \mathbf k ;\omega_q, \mathbf q)$ on $\mathbf k$ thus effectively reduces to the angle $\theta_{\mathbf q  \mathbf k} = \theta_{\mathbf q}-\theta_{\mathbf k} $ (i.e., the angle between ${\bf k}$ and ${\bf q}$), which we symbolically express by using the notation $f(\omega_k, \theta_{\mathbf q  \mathbf k} ;\omega_q, \mathbf q)$. In this language, Eq.~\eqref{eq:gdf} reduces to
\begin{align}
	\label{eq:gdf2}
	\int \frac{d\xi}{ 2 \, \pi } \, G^<(\omega_k ,\xi,\theta_\bk;t, \br )
	= i \, f(\omega_k ,\theta_\bk;t, \br )\,,\quad
	i  \int \frac{d\xi}{ 2 \, \pi }  \, G^>(\omega_k ,\xi,\theta_\bk;t, \br )
	= 1-f(\omega_k ,\theta_\bk;t, \br ) \,.
\end{align}
The idea is that, since the fermionic self-energy [cf. Eq.~\eqref{eq-ferm-self-en}] depends only on the frequency [viz. $\Sigma_0^R(\omega_k, \bk )\equiv \Sigma_0^R(\omega_k)$], $ A(\omega_k, \bk) \equiv A(\omega_k; \xi_\bk,\theta_{\bk})$ is a peaked function of $\xi \, \equiv \xi_\bk$ around $\xi=0$, when we change the variables from $\bk$ to its magnitude (parametrized by $\xi_\bk$ as the deviation from the equilibrium Fermi momentum) and angular orientation (encoded in $\theta_{\bk}$). Consequently, the structure of $G^{<,>}$ follows suit.

In the following subsections, we will consider two scenarios: (A) the bosons are assumed to be in equilibrium, with no bosonic kinetic equation as a consequence; (B) the bosons are slightly away from equilibrium, giving rise to coupled QBEs governing the dynamics of both the fermions and the bosons.

\subsection{Fermionic QBE when the critical Bosons are at equilibrium}
\label{sec-eqm-bos}

In this subsection, we consider the simplest scenario where we assume the bosons to be in equilibrium, such that
we can use the equilibrium form of the bosonic distribution function, $n_B (\omega) = ( e^{\beta \, \omega} - 1 )^{-1}  $. This also means that the bosonic Green’s functions do not depend on the relative coordinates, and we need to derive the QBE only for the fermions, which are assumed to be slightly away from the equilibrium. A review of the salient steps for deriving the fermionic part of the QBEs is provided in Appendix~\ref{appfer0}. Using Eqs.~\eqref{eqlhs} and \eqref{eqrhs} therein, we 
integrate over $ \omega_p $ on both sides of the QBE, such that it reduces to
\begin{align}
\label{final_eq}
& \left (  \Omega - v_F \,|\mathbf q| \cos{\theta_{\bp  \bq}}\right ) 
u(\theta_{\bp \bq};  \Omega, \bq )  +  I_1 = I_2\,,
\end{align} 
where
\begin{align}
& u( \theta_{\bp \bq};  \Omega, \bq )=
\int_{-\infty}^{\infty} \frac{d\omega }{ 2\, \pi  } 
\,  \delta f( \omega , \theta_{ \bp \bq};   \Omega, \bq )\,,\nn
& \frac{ I_1 } {e^2 \, N_0 } = 
\int \frac{d\theta_{\bp'\bq}}{ 2 \, \pi } \int_{-\infty}^{\infty}
\frac{d\omega'}{ 2 \, \pi } \int_{-\infty}^{\infty}
\frac{d\omega}{ 2 \, \pi } \, \text{Re} 
[ D^R_0 (\omega'-\omega, \bp-\bp') ]
\left [ \delta f( \omega, \theta_{\bp \bq})- \delta f(\omega', \theta_{\bp' \bq})\right] 
\left[ f_0 \bigg (\omega'  +  \frac{\Omega}{2} \bigg )
- f_0 \bigg (\omega'-\frac{\Omega}{2} \bigg )\right] ,\nn
& \frac{I_2} {  i \,e^2 \,N_0 }
\nn & =  \int \frac{d\theta_{\bp'\bq}}{ 2 \, \pi }
\int_0^\infty \frac{d\nu}{\pi} \,
\text{Im} [D_0^R( \nu, \bp-\bp')] \int_{-\infty}^\infty d\omega' 
\int_{-\infty}^\infty \frac{d\omega}{ 2 \, \pi } 
\left[ \delta f( \omega, \theta_{\bp\bq})-\delta f( \omega, \theta_{\bp'\bq})\right]
\Big[ \delta(\omega'-\omega  +  \nu) \left \lbrace 1 + n_B(\nu)-f_0(\omega') \right \rbrace
\nn  &  \hspace{ 12.5 cm } 
+  \delta(\omega'-\omega-\nu) 
\left \lbrace n_B(\nu) +  f_0(\omega') \right \rbrace  \Big ] \,.
\end{align}
We note that $u(\theta_{\bp \bq};  \Omega, \bq )$ is the physical quantity that quantifies the Fermi-surface displacement.

In this paper, we solely focus on the zero-temperature limit, in which the Fermi- and Bose-distribution functions reduce to $f_0(\omega)=\Theta(-\omega)$ and $ n_B(\omega)=0$, respectively. This leads to the simplified expressions of
\begin{align}
I_1  & = k_F \, v_F \int \frac{d\theta_{\bp'\bq}}{ 2 \, \pi } 
\left[u(\theta_{\bp\bq};  \Omega, \bq )-u(\theta_{\bp'\bq};  \Omega, \bq )\right]
F^R(\theta_{\bp\bp'},\Omega)\,, \nn
I_2  & = - \, i \, k_F \, v_F  \int \frac{d\theta_{\bp'\bq}}{ 2 \, \pi } 
\left[u(\theta_{\bp\bq};  \Omega, \bq )-u(\theta_{\bp'\bq};  \Omega, \bq )\right] 
F^I(\theta_{\bp\bp'},\Omega)\,,
\label{eqI1I2}
\end{align}
where
\begin{align}
\label{F-eq} 
F^R(\theta,\Omega)
& =\frac{\sin ({\theta} / {2} )}{\pi} 
\tan^{-1}\bigg(\frac{a}
{2 \sin^3 ({\theta} /{2})}  \bigg)  \,, \quad
F^I(\theta,\Omega)  =
\frac{\sin (|\theta|/2)} {2 \,\pi} 
\ln \Big(1 +  \frac{a^2} { \sin^6 (\theta/2)} \Big)\,,
\quad  a =\frac{\chi \,\Omega}{8 \,k_F^3}\,.
\end{align}
Thus, $F^R$ and $F^I$ define the real and imaginary parts of the GLF, $F=  F^R +  i \, \, F^I$. Hence, the final form of the QBE is obtained as
\begin{align}
& \left(\bar{\Omega}- \bbq \cos\theta\right) u(\theta ;\Omega,\bq )
+  \int \frac{d \tilde \theta} { 2\, \pi }  
\left[ u(\theta ;\Omega,\bq)- u( \tilde \theta ;\Omega,\bq ) \right] 
F(\theta-\tilde \theta,\bar{\Omega}) = 0\,, 
\nn & \bar{\Omega}= \frac{\Omega} {v_F \,k_F} \,, \quad
	\bbq =\frac{ |\bq| } { k_F} \,, \quad
	\theta \equiv\theta_{\bp \bq} \,, \quad
	\tilde \theta \equiv \theta_{\bp' \bq}\,,
	\label{eqQBE1}
\end{align}
and $u(\theta)\equiv u(\theta;\Omega,\bq)$. It is important to note that Eq.~\eqref{eqQBE1} is a self-consistent equation in the variables $\Omega$ and $\theta$.

In our earlier work \cite{ips-kazi}, we studied Eq.~\eqref{eqQBE1} in the collisionless limit such that the imaginary part of the GLF, $F^I$, was ignored. Since $F^I$ causes the damping of the collective modes, we assumed there that the decaying parts of the modes are negligible. But here, we explicitly include the effects of $F^I$, on the same footing as $ F^R $, to quantitatively estimate if they can genuinely be ignored or whether they destabilise the collective modes. 
In order to make the calculation analytically tractable, we decompose $F(\theta,\Omega)$ into angular-momentum channels, indexed by $\ell \in 0 \cup \mathbb{Z}^+$.

\subsection{Bosons away from the equilibrium}
\label{sec-neq-bos}

In this subsection, we remove the assumption that the bosons are in equilibrium. We allow both the fermionic and the bosonic modes to be slightly away from the equilibrium condition, with fluctuating nonequilibrium distribution functions, and study the corrections to the Fermi-surface's deformation modes.

Following the same approach as with the fermions (cf. Appendix~\ref{appfer}), the bosonic part of the QBEs is captured by
\begin{align}
&    \left[D_0^{-1}(p)-\text{Re} [\Pi^R ( p, x_{\rm rel})] , 
\, D^< ( p, x_{\rm rel}) \right]_{\text{PB}}- 
\left[\Pi^< ( p, x_{\rm rel}) ,\text{Re}[D^R ( p, x_{\rm rel})] \right]_{\text{PB}}
\nn & 
= - \, i  \left [ \Pi^> ( p, x_{\rm rel})  \, D^< ( p, x_{\rm rel})   
-\Pi^< ( p, x_{\rm rel})  \, D^> ( p, x_{\rm rel}) \right ].
\label{eqbosQBE1}
\end{align}
Since $\text{Re}[D^R] $ is a flat function of $\omega_p$, it can be safely ignored in Eq.~\eqref{eqbosQBE1}. We note that $\Pi^R$ is purely imaginary at equilibrium and, we assume that, slightly away from the equilibrium, it retains this property. With these approximations, the left-hand side of Eq.~\eqref{eqbosQBE1} simplifies to $\left[D_0^{-1}(p),
\, D^< ( k, x_{\rm rel}) \right]_{\text{PB}} = 2 \,\bp \cdot \nabla_{\br} D^<(\omega_p, \bp;t, \br )$. Transforming everything to the Fourier space, integrating over $\omega_p $ on both sides, and using Eq.~\eqref{eq-bos-dist}, we get
\begin{align}
2 \, \bp \cdot \nabla_{\br} n( \omega_p, \bp;t, \br ) = - \, i  
\left [ \Pi^>( \omega_p , \bp;t, \br )
\, n( \omega_p, \bp ;t, \br )-\Pi^<(\omega_p , \bp;t, \br )
\left \lbrace 1 + n( \omega_p, \bp ;t, \br ) \right \rbrace \right ].
\label{eqbosQBE3}
\end{align}
Next, we linearise Eq.~\eqref{eqbosQBE3} near equilibrium in the small deviations of $\delta n \equiv n-n_B$ and $\delta \Pi^{<(>)} \equiv \Pi^{<(>)}-\Pi_0^{<(>)}$, where $\Pi_0^{<}$ and $\Pi_0^{>}$ are the equilibrium self-energies and $n_B$ is the equilibrium value of the bosonic distribution function. Since $n_B(\omega)=0$ at $ T =0 $, we have
\begin{align}
2 \, \bp \cdot \nabla_{\br} n( \omega_p, \bp ;t, \br )  & = 
-\,i \left[\left \lbrace  \Pi_0^>(\omega_p, \bp)-\Pi_0^<(\omega_p , \bp) \right \rbrace
\delta n(\omega_p , \bp ;t, \br ) -\delta\Pi^<( \omega_p,\bp ;t, \br ) \right]
\nn
& = - 2\, \text{Im}[\Pi_0^R( \omega_p, \bp)]\,\delta n( \omega_p ,\bp ;t, \br ) 
+  i\, \delta\Pi^<( \omega_p ,\bp ;t, \br )\,.
\label{eqbosQBE4}
\end{align}
On Fourier-transforming from ($t, \br $) to ($  \Omega, \bq $), we get  
\begin{align}
\delta n( \omega_p , \bp ;  \Omega, \bq ) = \frac{1}{2}
\, \frac{\delta \Pi^<( \omega_p, \bp;  \Omega, \bq )}
{\bp \cdot \bq +  i \,  \text{Im}[\Pi_0^R( \omega_p ,\bp )] }\,.
\label{eqdelta_n}
\end{align} 
Up to the one-loop order,
\begin{align}
\Pi^<(x_1,x_2) &= -\,2\,i \, e^2 \, G^<(x_1,x_2) \, G^>(x_2,x_1)  \nn
\Rightarrow   \Pi^< ( \omega_p , \bp ;t, \br ) & = -\,2\, i \,e^2 
\int \frac{d^2\bk}{( 2 \, \pi )^2}    \int_{-\infty}^{\infty}\frac{d\nu}{ 2 \, \pi } 
\,G^<( \omega_p + \nu, \bk +\bp ;t, \br ) 
\, G^>( \nu , \bk ;t, \br ) \nn
& = -\,2 \,i \,e^2 \int \frac{d^2\bk}{( 2 \, \pi )^2}  \int \frac{d^2\bk'}{( 2 \, \pi )^2}  
\int_{-\infty}^{\infty}   \frac{d\nu}{ 2 \, \pi }
\int_{-\infty}^{\infty} \frac{d\nu'}{ 2 \, \pi } 
\,    G^<( \nu', \bk'; t, \br ) \, G^>( \nu ,\bk ;t, \br )
\,\delta(\bk'-\bp-\bk)\, \delta(\nu'-\omega_p-\nu) \nn
& = - \, 2 \, i \, e^2\, N_0^2 
\int_0^{ 2 \, \pi }\frac{d\theta_\bk}{ 2 \, \pi }
\int_0^{ 2 \, \pi }
\frac{d\theta_{\bk'}}{ 2 \, \pi }
\int_{-\infty}^{\infty}\frac{d\nu} { 2 \, \pi }
\int_{-\infty}^{\infty}\frac{d\nu'}{ 2 \, \pi } \,f( \nu' , \theta_{\bk'};t, \br ) \
\, [ \, 1-f( \nu ,\theta_\bk ;t, \br ) \, ]
\nn & \hspace{ 8 cm }
\times \delta(\theta_{\bk'}-\theta_\bp-\theta_\bk) \, \delta(\nu'-\omega_p-\nu)\,.
\label{pi equation 3}
\end{align}  
This leads to
\begin{align}
\frac{ \delta \Pi^<( \omega_p, \bp;t, \br ) } 
{ -\,2 \,i\, e^2 \, N_0^2 } =
\int_0^{ 2 \, \pi } \frac{d\theta_\bk}{ 2 \, \pi } 
\int_{-\infty}^{\infty}\frac{d\nu}{ 2 \, \pi } \,\delta f( \nu, \theta_{\bk}; t, \br )
\left [ 1-f_0(\nu-\omega_p)-f_0(\nu  +  \omega_p) \right ]. 
\label{eqpi4}
\end{align}
Plugging Eq.~\eqref{eqpi4} into Eq.~\eqref{eqdelta_n}, and using $\text{Im} [\Pi_0^R( \omega_p \bp)] = - \,\chi\, \omega_p/|\bp| \simeq \, - \,|\bp|^2$, we find that
\begin{align}
\delta n( \omega_p, \bp;  \Omega, \bq ) = \frac{ e^2 \, N_0^2 \int_0^{ 2 \, \pi } 
\,   \frac{d\theta_{\bk\bq}}{ 2 \, \pi } 
\int_{-\infty}^{\infty}\frac{d\nu}{ 2 \, \pi } \,
f( \nu, \theta_{\bk\bq} ;  \Omega, \bq )
\left [ \,1-f_0(\nu-\omega_p)-f_0(\nu  +  \omega_p) \, \right ]} 
{|\bp|^2  +  i \,\bp \cdot \bq} \,.
\end{align}

Now we are ready to compute the effects of $\delta n$ on $\delta f$, governed by Eqs.~\eqref{eq_int_qbe} and \eqref{eq-int-coll} (appearing in Appendix~\ref{appfer}). As for the fermionic QBE here, it takes the form,
\begin{align}
& (\Omega-v_F\, |\bq| \cos\theta_{\bp\bq}) \,\delta f ( \omega_p, \theta_{\bp \bq}; \Omega, \bq)
-\left[\text{Re} [\Sigma_0^R \Big (\omega_p  +  \frac{\Omega}{2} \Big )]
-\text{Re} [\Sigma_0^R \Big (\omega_p-\frac{\Omega}{2} \Big )] \right] 
\delta f  ( \omega_p, \theta_{\bp \bq}; \Omega, \bq )   
\nn &  
+  \left[ f_0 \Big(\omega_p  +  \frac{\Omega}{2} \Big)
- f_0 \Big (\omega_p-\frac{\Omega}{2} \Big ) \right] 
\delta  \text{Re} [\Sigma^R ( \omega_p , \theta_{\bp\bq};  \Omega, \bq )]  
\nn & = f_0(\omega_p) \, \delta\Sigma^>( \omega_p, \theta_{\bp\bq} ; \Omega, \bq )  
+  \Sigma_0^>(\omega_p)
\,\delta f( \omega_p, \theta_{\bp\bq} ;  \Omega, \bq )- \left[ f_0(\omega_p)-1 \right ]
\, \delta \Sigma^<( \omega_p, \theta_{\bp\bq} ;  \Omega, \bq ) 
-\Sigma_0^<(\omega_p) \, \delta f ( \omega_p , \theta_{\bp \bq}; \Omega, \bq) \,.
\label{eq-ferm-fin}
\end{align} 
The corrections due to the nonequilibrium bosons appear through the fermionic self-energy terms, viz. $\delta \Sigma^>$, $\delta \Sigma^< $, and $ \delta \text{Re} [ \Sigma^R] $, which are now functions of both $\delta n$ and $\delta f$. We will denote the contribution coming from $\delta n$ as $\delta \Sigma_{fb}^{>(<)}$, which is equal to $\delta \Sigma^{>(<)}
-\delta \Sigma_{b0}^{>(<)} $, where $\delta \Sigma_{b0}^{>(<)} $ is the contribution when the bosons are in equilibrium.  A straightforward calculation, using Eqs.~\eqref{sigma less in equilibrium}, \eqref{sigma great in equilibrium}, and \eqref{eqre-self-en-eqm}, shows that
\begin{align} 
&   \delta \Sigma_{fb}^>(\omega_p, \theta_{\bp\bq} ;   \Omega, \bq ) =
i \, e^2 \, N_0 \int_0^{ 2 \, \pi } \frac{d\theta_{\bk\bq}}{ 2 \, \pi }
\int_0^{\infty}\frac{d\omega}{\pi} 
\,   \text{Im} [D_0^R( \omega, \theta_{\bk\bp} )]\, 
\delta n( \omega, \theta_{\bk\bp};  \Omega, \bq )
\left[f_0(\omega_p  +  \omega) + f_0(\omega_p-\omega)-2 \right],
\nn &    \delta \Sigma_{fb}^<(\omega_p, \theta_{\bp\bq} ;   \Omega, \bq ) =
i \, e^2 \, N_0 \int_0^{ 2 \, \pi } 
\frac{d\theta_{\bk\bq}}{ 2 \, \pi }\int_0^{\infty}
\frac{d\omega}{\pi} \, \text{Im} [D_0^R( \omega, \theta_{\bk\bp}) ]\, 
\delta n( \omega, \theta_{\bk\bp} ;  \Omega, \bq )
\left[ f_0(\omega_p  +  \omega) + f_0(\omega_p-\omega) \right],
\nn & \delta \text{Re}[\Sigma_{fb}^R(\omega_p, \theta_{\bp\bq} ;   \Omega, \bq )] = 0 \,.   
\label{eq_sig_gr_less}
\end{align}
We first observe that, since there is no correction to $\delta \text{Re} [\Sigma^R] $ when one includes the boson dynamics, the left-hand side Eq.~\eqref{eq-ferm-fin} remains unchanged. On the other hand, the correction to the collision integral due to the bosonic density fluctuations is captured by
\begin{align}
\delta I_{fb} & =  
f_0(\omega_p) \, \delta\Sigma_{fb}^>(\omega_p, \theta_{\bp\bq} ;   \Omega, \bq )
- \left [\, f_0(\omega_p)-1 \,\right ]
\delta \Sigma_{fb}^<(\omega_p, \theta_{\bp\bq} ;   \Omega, \bq ) \nn
&= -i \,e^2 \,N_0 \int_0^{2\pi} \frac{d\theta_{\bk\bq}}{2 \,\pi}
\int_0^{\infty}\dfrac{d\omega}{\pi}  
\, \text{Im} [ D_0^R( \omega, \theta_{\bk\bp} )] \,
\delta n( \omega, \theta_{\bk\bp};  \Omega, \bq ) 
\left[f_0(\omega_p + \omega) + f_0(\omega_p-\omega)-2 \,  f_0(\omega_p)\right].
\label{eq-coll-int}
\end{align}
Even though $\delta I_{\text{fb}}$ is finite, and it couples $\delta f$ and $\delta n$ self-consistently, we find that the Fermi surface displacement [viz. $u(\theta_{\bp};  \Omega, \bq )$], as arises in Eq.~\eqref{final_eq}, does not depend on $\delta n$. This can be seen by integrating $\delta I_{\text{fb}}$ over the fermionic frequency, $\omega_p$. Because the integrand is an odd function of $\omega_p$, we have
\begin{align}
	\int_{-\infty}^\infty \frac{d\omega}{ 2 \, \pi }
	\, \delta I_{\text{fb}}\propto  \int_{-\infty}^\infty \frac{d\omega}{ 2 \, \pi } \left[f_0(\omega_p  +  \omega)
	+  f_0(\omega_p-\omega)-2 f_0(\omega_p)\right]=0\,.
\end{align}
Hence, we conclude that we end up with the same collective mode equation [viz. Eq.~\eqref{final_eq}], without any correction coming from the fluctuating bosonic density.

\section{Explicit solutions in various angular-momentum channels}
\label{secsol}

In this section, we will explicitly work out the solutions to Eq.~\eqref{final_eq}. It is helpful to understand the nature of the collective modes if we first focus on the $\ell  = 0$ modes and eventually take upon the task of solving for nonzero $\ell$-values as well. The two subsections are devoted to these two scenarios.

\begin{figure*}[t!]
	\includegraphics[width=0.35 \textwidth]{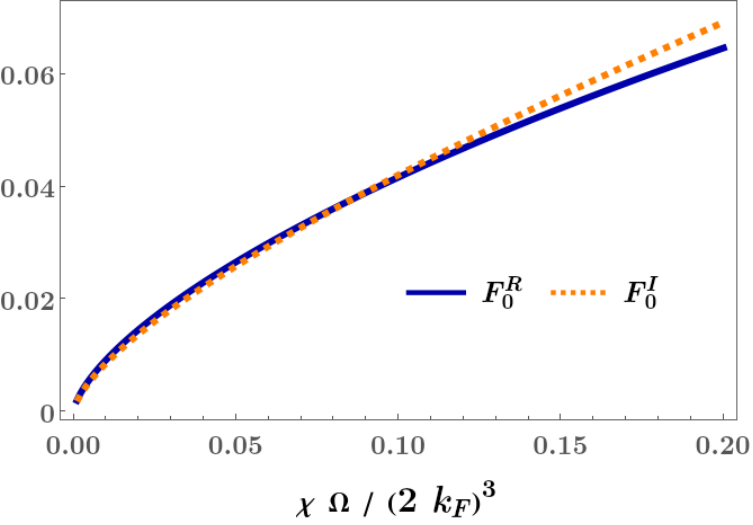}
	\caption{Frequency dependence of the generalised Landau parameter in the $\ell =0$ angular-momentum component of $ F(\theta,\Omega)$ [cf. Eq.~\eqref{eqF0}], after resolving it into the real and imaginary parts.\label{figLF}}
\end{figure*}

\subsection{Zero angular-momentum channel: $F_0$ model}
\label{seczero}

In this subsection, we focus on the dominant behaviour, which can be extracted from the $\ell = 0$ component of $F(\theta,\Omega)$, captured by
\begin{align}
\label{eqF0} 
F_0(\Omega)=\int_0^{ 2 \, \pi }\frac{d\theta}{ 2 \, \pi } \, F(\theta,\Omega)\,.
\end{align}
We approximately solve Eq.~\eqref{eqQBE1} setting $F(\theta,\Omega) \simeq F_0(\Omega)$, calling it the ``$F_0$ model''. In Fig.~\ref{figLF} we illustrate the frequency dependence of $F_0^R \equiv \text{Re} [ F_0] $ and $F_0^I \equiv 
\text{Im} [F_0 ] $ via the function $a = \bOmega \, \alpha.$ 
In the low-frequency limit of $\bOmega \, \alpha \ll 1$, both $F_0^R$ and $F_0^I$ are of the same order and behave in a qualitatively similar way. Explicitly, their forms are $F_0^R(\bOmega) 
\approx 0.4 \,(\alpha \, \bOmega)^{ \frac{2} {3} }$ and $F_0^I (\bOmega)\approx 0.36\, (\alpha \,\bOmega)^{ \frac{2} {3} } $.
In the region when they are approximately equal, we get
\begin{align}
\label{eqf02}
F_0^R(\bOmega) = F_0^I(\bOmega)=\Omega_0^{ \frac{1} {3} } \, \bOmega^{ \frac{2} {3} }\,,
\text{ where }    \Omega_0 = 0.06 \, \alpha^2 =  \frac{ 0.0009 \,e^4} { 4 \,\pi^2  \, v_F^2 \,k_F^2}\,.
\end{align} 
This simplifies Eq.~\eqref{eqQBE1} to
\begin{align}
\frac{1}{F_0(\bOmega)}=\int_0^{ 2 \, \pi }\frac{d\theta}{ 2 \, \pi } \,
\frac{1}{\bar{\Omega}-\bbq \cos\theta + F_0(\bOmega)} =
-\frac{2}{\bbq}\oint \frac{dz}{ 2 \, \pi  i} \,\frac{1}{(z-z_ + )(z-z_-)} \,,
\label{eq-self-con}
\end{align}
on changing to the complex variables, $z_\pm=s\pm \sqrt{s^2-1}$, with $s=(\bOmega + F_0(\bOmega))/\bbq$.

Let us first discuss our conclusions reported in our earlier work, viz. Ref.~\cite{ips-kazi}.
When $s$ is real, which is the case considered in Ref.~\cite{ips-kazi} where we set $F_0^I=0$, the right-hand side of Eq.~\eqref{eq-self-con} gives $1/\sqrt{s^2-1}$ when $s>1$, and the dispersion relation is then governed by
\begin{align}
	\bOmega^2\left[1 +  2 \left(\frac{\Omega_0}{\bOmega}\right)^{ \frac{1} {3} }\right] = \bbq^2\,.
	\label{eq_no_damp}
\end{align}
Eq.~\eqref{eq_no_damp} indicates that $\bOmega \sim \Omega_0$ defines a \textit{crossover scale} --- on moving across it, the dispersion acquires contrasting asymptotes. More explicitly, for $\Omega \ll \Omega_0$, one can ignore the constant term in the parenthesis and the dispersion relation reduces to $\Omega \propto \bbq^{6/5}$. On the other hand, for $\Omega \gg \Omega_0$, dispersion is linear  in momentum: $\bOmega \propto \bbq$. For $s\leq 1$, the right-hand side of Eq.~\eqref{eq-self-con} vanishes. Our analysis in Ref.~\cite{ips-kazi} shows that this regime represents the particle-hole excitations, just like a Fermi liquid, and their dispersions form an energy band. The boundary for the dispersion of this continuum of excitations is given by the condition $ s = 1 \Rightarrow \bar{\Omega}+ F^R_0(a) = |\bar{\mathbf q}| $, and changes qualitatively across $\Omega_0$, such that the curve goes as (a) $  \bar \Omega \sim |{\bar \bq}|^{3/2}$ for $   \Omega < \Omega_0$, and (b) $  \bar \Omega \sim |\bar \bq| $ for ${\Omega} > {\Omega_0}$. These were the key findings of Ref.~\cite{ips-kazi}.

\begin{figure}[t!]
	\subfigure[]{\includegraphics[width = 0.32 \textwidth]{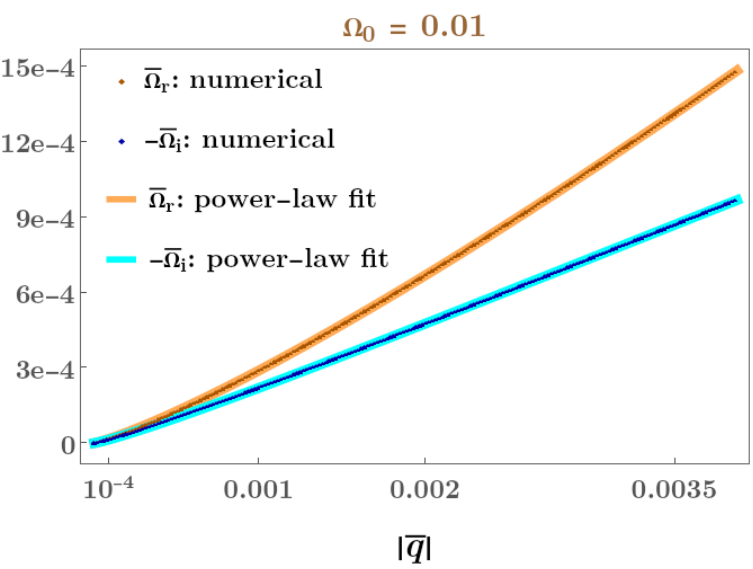}}\hspace{0.35 cm}
	\subfigure[]{\includegraphics[width = 0.315 \textwidth]{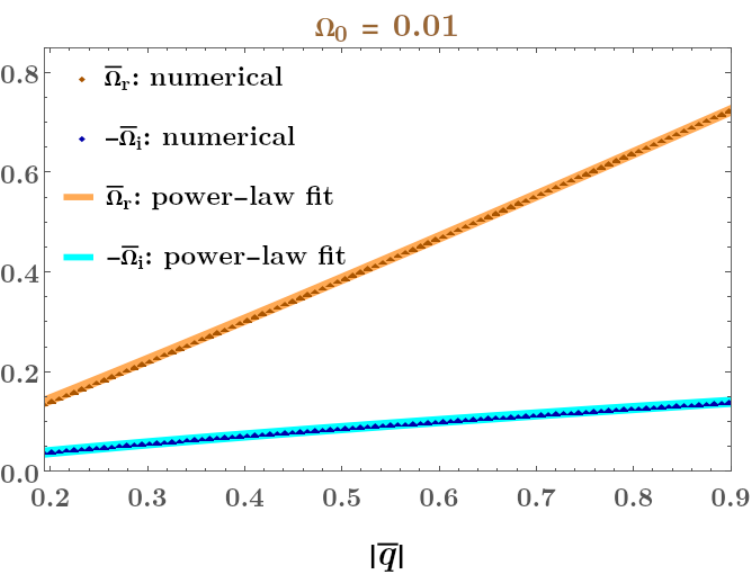}}
	\subfigure[]{\includegraphics[width = 0.33 \textwidth]{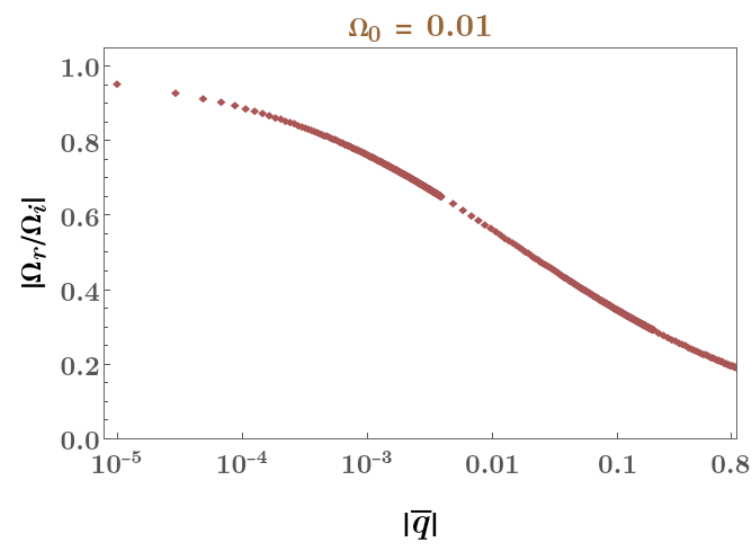}}
	\caption{Dispersion relation of the zero-sound mode, showing the behaviour of the real and imaginary parts, $\bOmega_r$ and $\bOmega_i $.  In subfigures (a) and (b), we have compared the numerical solutions of Eqs.~\eqref{eq-dis-damp} and \eqref{imaginary part} with the analytically-estimated power-law fittings, depending on the energy scales, as discussed in the text. Subfigure (c) shows that the ratio of $|\Omega_i|$ and $|\Omega_r|$ (obtained from the numerical solutions) falls off rapidly with increasing $\bbq$.\label{fig:dispersion}}
\end{figure}

In the presence of the $F_0^I$ term, the collective-mode frequency can be parametrised as $\Omega = \Omega_r +  i \, \, \Omega_i$, where a nonzero imaginary part, $ i\, \Omega_i$, is explicitly written out. This implies that we must have, in general, $s = s_r + i\, s_i$, where there a nonzero imaginary part in $s$ as well. In fact, $s_r = (\bOmega_r + F_0^R)/\bbq$ and $s_i=(\bOmega_i + F_0^I)/\bbq$. Let us focus on the positive values of the real part (i.e.,  $s_r\geq 0$), for which $|z_-|<1$, resulting in
\begin{align}
	F_0^R(\bOmega) +  i \, \, F_0^I(\bOmega) = \bbq \; \sqrt{(s_r +  i \, \, s_i)^2-1} \,.
	\label{eq-self-con2}
\end{align}
Separating the real and imaginary parts of Eq.~\eqref{eq-self-con2}, we get two real-valued equations:
\begin{align}
	\tilde{s}_r^2-\tilde{s}_i^2-\bbq^2  & = F_0^R(\bOmega)^2-F_0^I(\bOmega)^2
	\text{ and }     
	\tilde{s}_r  \, \tilde{s}_i   = F_0^R(\bOmega)\, F_0^I(\bOmega) \,,
	\text{ where }     \tilde{s}_{r} = \bOmega_{r} + F_0^{R}(\bOmega)
	\text{ and } \tilde{s}_{i} = \bOmega_{i} + F_0^{I}(\bOmega)\,.
	\label{eq-s-re-im}
\end{align}
Plugging in $\tilde{s}_i =
F_0^R \, F_0^I/\tilde{s}_r$, the above equations reduce to
\begin{align}
	{\widetilde \Omega_r}^2 \left[ 1 
	+ 2 \, \widetilde \Omega_r^{ - \frac{1} {3} }\right]
	\left[1 + 2 \,{\widetilde \Omega_r}^{ -\frac{1} {3} }  
	+  2 \,{ \widetilde \Omega_r}^{ - \frac{2} {3} } \right] =
	\left[1  +  {\widetilde \Omega_r}^{ - \frac{1} {3} }
	\right]^2 {\tilde q}^2 \,,
	\text{ where }
	\widetilde \Omega_r = \frac {\bOmega_r} {\Omega_0} \text{ and }
	\tilde q = \frac { \bbq } {\Omega_0}\,.
	\label{eq-dis-damp}
\end{align}
Comparing Eq.~\eqref{eq-dis-damp} with Eq.~\eqref{eq_no_damp}, we find that the effect of not throwing away $F_0^I$ is manifested in extra terms. 
Additionally, we have a nonzero imaginary part behaving as
\begin{align}
	\bOmega_i = \frac{ [ F_0^R(\bOmega_r) ]^2 }
		{\bOmega_r + F_0^R(\bOmega_r)}-F_0^R(\bOmega_r)\,.
		\label{imaginary part}
	\end{align}
	We can determine the asymptotic behaviour in two opposite limits as follows:
	\\(1) For $ {\widetilde \Omega_r} \ll 1$, the solutions are obtained as $ {\widetilde \Omega_r} \simeq
	(\tilde q / 2)^{6/5}
	\Rightarrow \bOmega_r \simeq \Omega_0^{-1/5} \left( \frac{\bbq } {2} \right)^{6/5}$ and
	$  \bOmega_i  \simeq 
	- \,\bOmega_r +  \Omega_0^{-3/5}
	\left( \frac{\bbq } {2} \right)^{8/5} $. In this regime, the applicable range of values of $\bbq $ is much less than $\Omega_0$. We have kept the subleading term (in terms of powers of $\bbq$ for $\bbq \ll 1$) in ${\widetilde \Omega_r}$'s expression because the coefficients are comparable and are essential to get a good fit with the full numerical solutions. In fact, the second term gives us the deviation of $|\bOmega_i|$ from $ | \bOmega_r |$, which increases monotonically with $\bbq$.
\\(2) For $ {\widetilde \Omega_r} \gg 1$, the solutions are obtained as $\bOmega_r \simeq  \bbq$ and $\bOmega_i \simeq
	-\,\Omega_0^{1/3} \,\bbq^{2/3} + \Omega_0^{2/3} \, \bbq^{1/3}$. In this regime, the applicable range of values of $\bbq $ is much greater than $\Omega_0$. Again, we have kept the subleading term (this time, in terms of powers of $\Omega_0$ for $ \Omega_0 \ll 1$) in ${\widetilde \Omega_r}$'s expression in order to get a good fit with the full numerical solutions.
In Fig.~\ref{fig:dispersion}, we show a comparison of the numerically-obtained solutions [of Eqs.~\eqref{eq-dis-damp} and \eqref{imaginary part}] with our analytical approximation of the power-law behaviour. Our solutions demonstrate the fact that the zero sound is indeed long-lived, because the magnitude of the imaginary part, $\bOmega_i$, is parametrically smaller than that of the real part, $\bOmega_r $. In fact, $|\Omega_i/ \Omega_r|$ falls off swiftly with $\bbq$, as illustrated in Fig.~\ref{fig:dispersion}(c). All these observations then uphold the conclusions of Ref.~\cite{ips-kazi}, where we had ignored $F^I$ and, hence, any measure of decay of the collective mode. We would like to emphasise that, there too, the dispersion relation was found to behave as $ \sim \bbq^{6/5}$ and $\sim \bbq $, in the regimes below and above a characteristic crossover scale, respectively.

\begin{figure}[t!]
\subfigure[]{\includegraphics[width = 0.35 \textwidth]{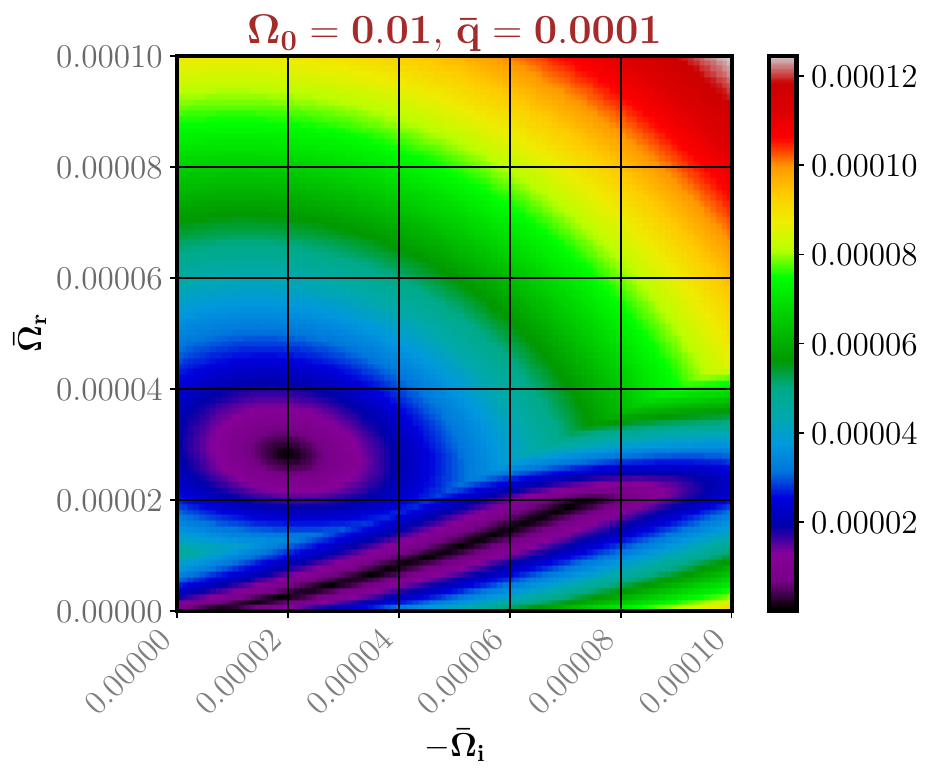}} \hspace{1 cm}
\subfigure[]{\includegraphics[width = 0.35 \textwidth]{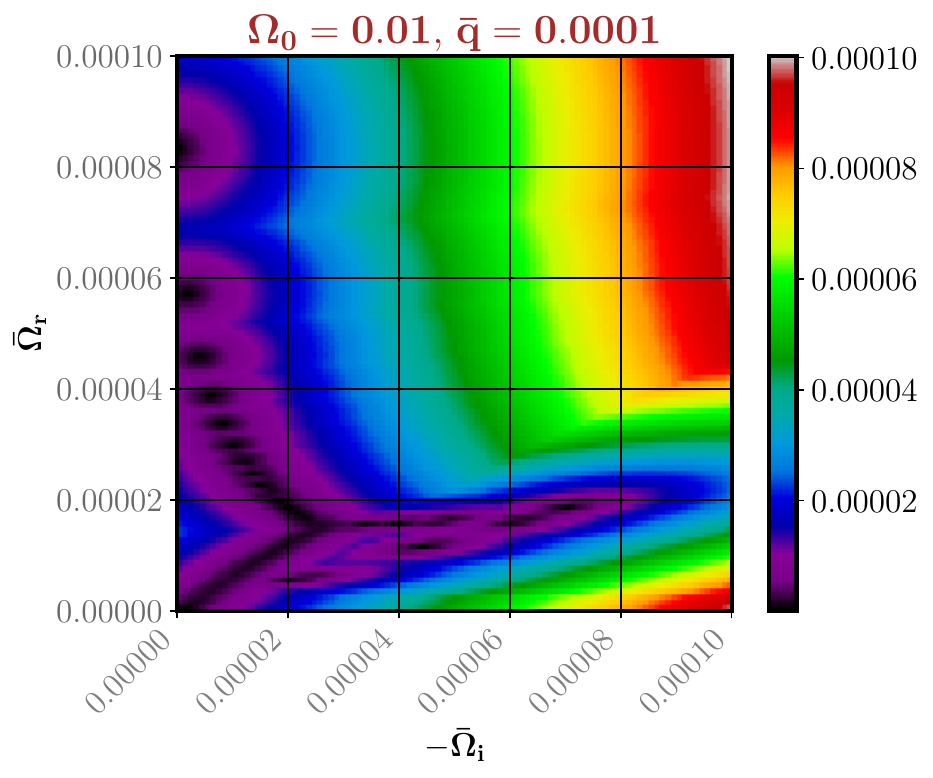}}
\subfigure[]{\includegraphics[width = 0.35 \textwidth]{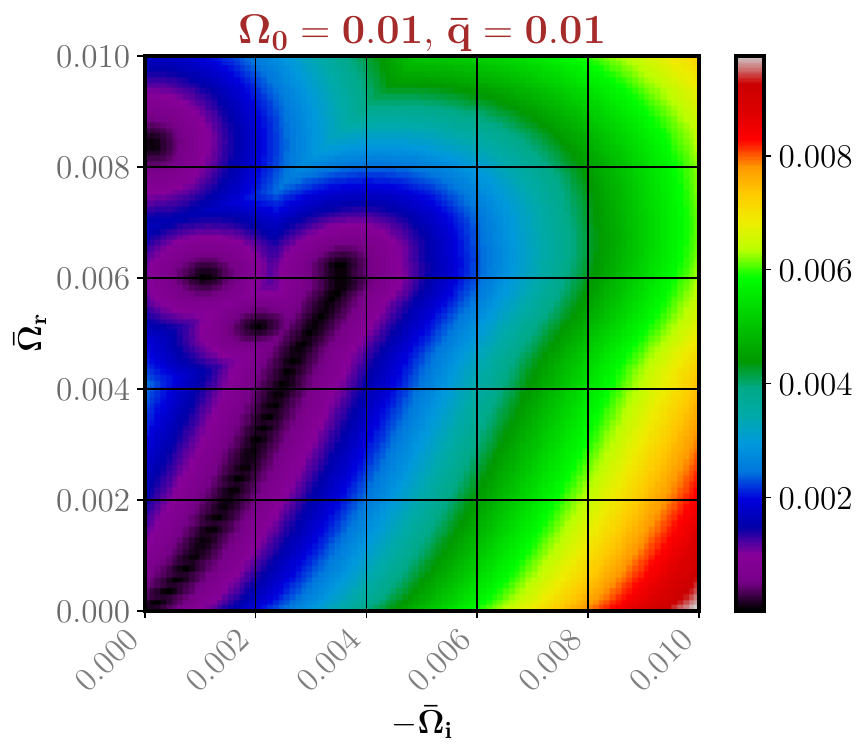}} \hspace{1 cm}
\subfigure[]{\includegraphics[width = 0.35 \textwidth]{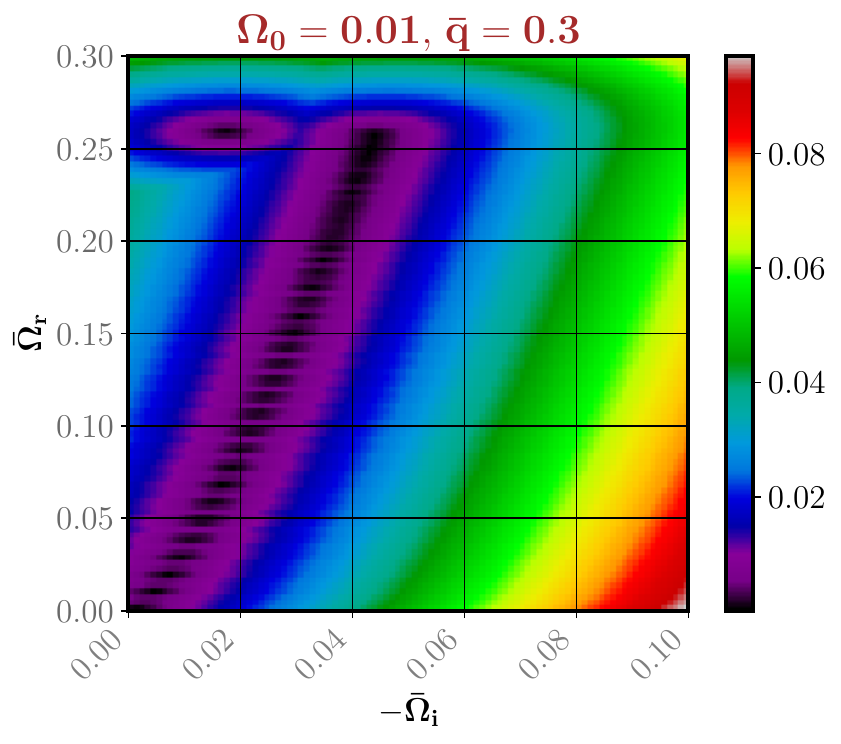}}
\caption{Contourplots of the magnitude of the lowest eigenvalue of $M$ [cf. Eq.~\eqref{eqM}], for some values of $\left ( \Omega_0, \bar q \right )$, in the complex plane of $\bar \Omega$. While subfigure (a) corresponds to the $F_0$-model of Sec.~\ref{seczero}, subfigures (b)-(d) capture the results for the $F_\ell$ model of Sec.~\ref{seccode}.\label{figevalue}}
\end{figure}


\subsection{Generic angular-momentum channels: $F_\ell$ model}
\label{seccode}

In this subsection, we include the effects of $\ell >0$ modes of $F(\theta)$ as well, which provides a full description of all the collective modes arising from Eq.~\eqref{eqQBE1}. Going beyond the analysis in Sec.~\ref{seczero}, where we approximated $F(\theta,\Omega)$ by retaining only the $\ell = 0$ mode, $F_0(\Omega)$, here we keep the nonzero $\ell$-components upto arbitrary order.  This needs to be solved numerically by truncating at some high value of $\ell$. In our earlier work \cite{ips-kazi}, where $F^I$ was ignored, we ended up with two distinctly different types of excitations: particle-hole like localised modes forming an energy band (or continuum) and delocalised collective modes with discrete energy-levels. Here, we incorporate the $F^I$-part appropriately, instead of setting it to zero, and investigate the fate of the collective modes in an exact way.

The expression in Eq.~\eqref{eqQBE1} is decomposed into angular momentum components by defining
\begin{align}
\label{eq_decomp}
	& u(\theta) =  u_0+2 \, \sum_{\ell=1} ^\infty  u_{\ell} \cos(\ell \, \theta )\,,
\quad
F (\theta,\Omega) = F_0(\Omega) + 2  \,\sum_{\ell=1} ^\infty  F_\ell (\Omega) \cos(\ell \, \theta )\,,
\quad F_{\ell}(\Omega) =  F_{\ell}^R(\Omega) + i \, F_{\ell}^I(\Omega)\,.
\end{align}
Here, $ u_{\ell}$ and $F_\ell({\Omega})$ are the amplitudes of $u(\theta)$ and $F(\theta,\Omega)$ in the $\ell$-channels, respectively. Since the fermion-fermion interactions depend only on the magnitude of the momentum-exchange encoded via $F(\theta,\Omega)$, they must comprise only even-parity configurations obeying $\theta \rightarrow  -\theta $. That motivates us to study the even-parity modes of $u(\theta)$.\footnote{We defer from providing an analysis of the odd-parity deformations [viz., $u(-\theta)=-u(\theta)$] as they are subleading compared to the even ones, and give rise to no collective-mode solutions~\cite{ips-kazi}.} Plugging in the expressions from Eqs.~\eqref{eq_decomp} into Eq.~\eqref{eqQBE1}, we find that the mother QBE decomposes into a set of recursive equations between different $\ell$ components, as shown below:
\begin{align}
\label{zero_mode_even}  
& \bar{\Omega}\,u_0-\bar{q} \,u_1 = 0 \,,\quad
u_{\ell}-  \dfrac{ 2\,\bar{\Omega}}{\bar{q}}
\left(1+\dfrac{F_0(\bar{\Omega})-F_{\ell-1}(\bar{\Omega})} {\bar{\Omega}}\right) u_{\ell-1}
+ u_{\ell-2} = 0 \text{ for } \ell \geq 2 \text{ and } \bar{q} \equiv |\bar{\mathbf q}|.
\end{align} 
One can easily identify that these equations are identical in form to those characterising a one-dimensional non-Hermitian tight-binding model, if we interpret the $\ell$-indices as playing the role of the lattice-site labels. In order to obtain the solutions, we need to solve for the coupled equations numerically by truncating the equations at some reasonable finite value of $\ell$, which we indicate by $N-1$.

\begin{figure}[t!]
\includegraphics[width = 0.75 \textwidth]{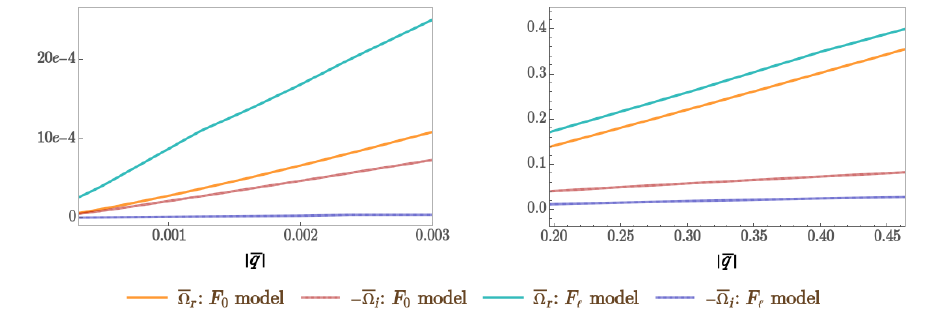}
\caption{Comparing the dispersion relation of the zero-sound mode, capturing the behaviour of both the real and imaginary parts ($\bOmega_r$ and $\bOmega_i $), calculated from the $F_0$ and $F_\ell$ models. While the left panel shows the extreme small-$|\bar {\mathbf q}|$ regime, the right one corresponds to somewhat larger-$|\bar {\mathbf q}|$ regime.\label{fig:zero_sound_disp}}
\end{figure} 

For notational simplification, we cast Eq.~\eqref{zero_mode_even} as a matrix equation, 
$	M(\bar{\Omega},\bar{q}) \; U=0$, where $U=\left(u_0,u_1,\cdots u_{N-1}\right)^T$ and
\begin{align}
	\label{eqM}
	M(\bar{\Omega},\bar{q}) = \begin{pmatrix}
		\bar{\Omega} & -\bar{q}  & 0 & 0 & 0 & \cdots \\
		-\bar{q}/2 & \bar{\Omega} + F_0 -F_1 & -\bar{q}/2 & 0 & 0 & \cdots\\
		0  & -\bar{q}/2 & \bar{\Omega} + F_0 -F_2 & -\bar{q}/2 & 0 & \cdots \\
		\vdots & \vdots & \vdots & \vdots & \vdots & \ddots 
	\end{pmatrix}_{N \times N} \,.
\end{align}
Because $\bar{\Omega} \equiv \bar{\Omega}_r+i\, \bar{\Omega}_i$ and $F_\ell \equiv F^R_\ell (\bar{\Omega}_r)+i \,F^I_\ell (\bar{\Omega}_r)$ are complex, $M$ itself is an $N\times N$-dimensional complex matrix.
The condition of $ M(\bar{\Omega},\bar{q}) \; U=0$ reduces to $\det  [M(\bar{\Omega},\bar{q})]=0$. This is equivalent to demanding that at least one of the eigenvalues of $M(\bar{\Omega},\bar{q})$ vanishes, which gives us the dispersion relation for a bonafide collective mode. We numerically find the eigenvalues of $M$ as functions of $\bar{\Omega}$ and check if any of them has zero value. In Fig.~\ref{figevalue}, we show the density-plots for magnitude of the  eigenvalues in the $\Omega_r \Omega_i$-plane for some fixed values of $\bar{q}$. The black regions are where the magnitude is the smallest (within the tolerance of our numerical precision), thus representing the existence of the physical modes.  In order to understand what new effects are brought about by the $\ell>0$ modes of $F(\theta)$, compared to the $F_0$ model, in Fig.~\ref{figevalue}(a) we show the density-plot exclusively for nonzero $F_0 $ only (i.e., by setting $F_{\ell>0} = 0 $ in $M$). Subfigures (b)-(d) illustrate the results for our generic-$F_\ell$ analysis, by setting $\bar{q}$ to $ 10^{-4}$, $0.01$, and $0.3$, respectively. We have used $ N= 80$ for all our numerical simulations.

In agreement with our earlier findings in Ref.~\cite{ips-kazi}, we identify two kinds of excitations: (1) An isolated region, analogous to the zero-sound solution of a Fermi liquid, in the regime labelled as $ \Omega = \Omega_{zs} $. We will designate it as the zero sound of the critical Fermi surface of the NFL. (2) A continuum region which can be interpreted as the particle-hole continuum, labelled as $\Omega = \Omega_{ph} $.
We find that, on including generic $\ell$ components, many new isolated regions develop, in addition to the zero sound [cf. Fig.~\ref{figevalue}(b)-(d)]. With decreasing momentum-values, the number ($n$) of these extra modes increases, and $ n \rightarrow \infty $ in the limit $\bq\rightarrow 0$, connecting the particle-hole continuum to the zero sound solution continuously. This is a qualitatively different scenario not captured by the drastic simplification of the $F_0$ model. Because of the appearance of these new solutions, the zero-sound solution (which is the farthermost island measured from the extended regions of the particle-hole continuum in the density-plots) is pushed more towards the $\bar{\Omega}_i = 0$ axis --- this shows that, in reality, the zero sound is much longer-lived compared to what is found within the $F_0$ model [cf. Fig.~\ref{fig:dispersion}]. We track the position of the center of the zero-sound island in the $\bar \Omega $-plane, as a function of $\bar q $, and extract the dispersion relation of the zero-sound mode in Fig.~\ref{fig:zero_sound_disp}.
Comparing its behaviour with the results from the $F_0$ model, we find that the discrepancy grows stronger and stronger as we move towards the lower and lower values of the momentum. Intriguingly, in the $ \mathbf q \rightarrow 0$ limit, $\Omega_i(\bq)\approx 0 $, while $\Omega_r(\bq)\propto |\bq| $ (by using a fitting function, we find that $ \Omega_r(\bq)\sim |\mathbf q|^{0.96}$). Therefore, in the extreme low-$|\mathbf q|$ limit, the scalings differ from those of the $F_0$-model (discussed in Sec.~\ref{seczero}), as visualised explicitly in the left panel of Fig.~\ref{fig:zero_sound_disp}. At somewhat higher momenta, the qualitative behaviour becomes similar, as observed in the right panel of Fig.~\ref{fig:zero_sound_disp}.

\begin{figure}[t!]
\includegraphics[width = 0.27\textwidth]{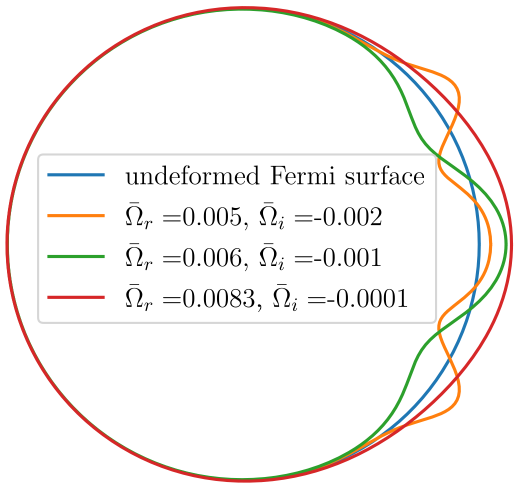}
\caption{Schematic plot of the deformations of the Fermi surface arising from the collective-mode solutions of the three tiny islands visible in Fig.~\ref{figevalue}(c), corresponding to $\bar q  = 0.01$. The $\bar \Omega$-values of the centres of these islands are shown in the plot-legends. The blue circle represents the undeformed Fermi surface.\label{figfermi}}
\end{figure}

In order to characterise the new discrete modes corresponding to the extra islands, we label the $\Omega$-values of their centres as $\Omega_\zeta (\bq)$. We have computed (numerically) the corresponding eigenvector, $U^\zeta$, satisfying $M(\bar{\Omega}_\zeta ,\bar q) \; U^\zeta = 0$. Defining $ U^\zeta_{\ell+1}$ as the $(\ell+1)^{\rm th}$-component of $U^\zeta $, the Fermi-surface displacement can be expressed as $u(\theta ) = u^\zeta (\theta)$, where
\begin{align}
u^\zeta (\theta) = \frac{1} {\mathcal N}
\left [ U^\zeta_0 + 2 \sum_{\ell>0} U^\zeta_\ell \cos (\ell \theta) \right ]
\text{ and }
\mathcal N  \equiv \sum_{\ell}|U^\zeta_\ell|^2
\end{align}
is the factor for normalizing $U^\zeta$. Fig.~\ref{figfermi} illustrates the deformations pictorially for the three islands (so that $\zeta \in \lbrace 1, 2, 3 \rbrace $) of Fig.~\ref{figevalue}(c) corresponding to $ \bar q =0.01 $. The red curve represents the zero sound (the island farthermost from the $\bar \Omega_r = 0$ axis), while the green and orange curves represent the modes due to the remaining two islands. Clearly, the latter represent higher-order harmonics of the net Fermi-surface deformation.

\section{Discussions, summary, and future outlook}
\label{secsum}
	
In this paper, we have analysed the nature of generic displacements of the critical Fermi surface for the NFL system arising at the Ising-nematic QCP, by using the quantum-Boltzmann-equation formalism applicable for our NFL scenario. By incorporating a generalised fermionic distribution function, we have derived the self-consistent equations whose solutions give us the functional forms of the Fermi-surface displacements as well as their dispersions. In particular, we have focused on determining the dispersion and stability of the low-energy and low-momentum excitations/collective modes, arising from these displacements. Compared with our earlier work, we have included here the effects of damping, originating from the collision integral for the fermionic QBE. We have discussed the implications in two scenarios: (1) The simplistic $F_0$ model, where we are able to find an analytical solution by simply setting $F(\theta,\Omega)=F_0(\Omega)$, retaining only the $\ell=0$ angular-momentum component. (2) The $F_\ell$ model, where generic $\ell\neq 0$ components are included, which warrants a numerical simulation to obtain the explicit results.  A summary of our findings is elucidated below:
\begin{enumerate}
	
\item Within the $F_0$ model, the dispersion of the zero sound of our NFL exhibits a $\Omega_r \sim |\mathbf q|^{6/5}$ behaviour at the lowest-energy scales, unlike the $\Omega_r \sim |\mathbf q|$ dependence predicted from the Fermi-liquid theory. This fractional-power-law dependence is seen to morph into an $ \Omega_r \sim |\mathbf q|$ behaviour when the energy is cranked up above a crossover scale, $\Omega_0$. As a consequence of accounting for the collision terms, the dispersion acquires a nonzero imaginary part, $\Omega_i $, which corresponds to damping. However, $|\Omega_i|$ is parametrically much smaller than $\Omega_r$, making the $\ell = 0$ mode a long-lived excitation. 
	$\Omega_i$ is also seen to cross over to distinct power-law characteristics with respect to $\Omega_0$.
	As a function of the mode's momentum, the ratio $|\Omega_i /\Omega_r |$ decays rapidly in a monotonic way (in all the applicable energy ranges), and becomes more and more negligible.
	
\item Incorporating the $F_\ell$ model, for very small momenta, the nature of the zero sound is found to have linear dependence on momentum and a negligible imaginary part. Intriguingly, in addition to the zero-sound solution, many new collective-mode excitations of the critical Fermi surface emerge, whose $\Omega $-values lie between those of the zero sound and particle-hole continuum (when visualised in the complex $\Omega$-plane). At extremely-low momenta, the number of these additional modes proliferate [cf. Fig.~\ref{figevalue}(b)]. However, with increasing $\bar q$, the extra islands disappear progressively and the overall behaviour crosses over to the results obtained from the $F_0$ model. 
	
\end{enumerate}  
We have extended our calculations to include the effects of the bosonic density fluctuations as well. Interestingly, the $T=0$ results for the Fermi-surface displacements remain unchanged under this generalisation.

Let us now compare our studies with other relevant studies related to NFLs:
\begin{enumerate}
\item Relation to earlier Boltzmann and kinetic approaches:
\\Traditional Boltzmann or kinetic-equation treatments of quantum critical metals typically rely either on the existence of long-lived quasiparticles or focus on the collisionless regime. In contrast, our analysis formulates quantum Boltzmann equations directly in terms of Keldysh Green’s functions and does not assume quasiparticle-coherence in the frequency space \cite{prange, kim_qbe, ips-zero-mode}. The key new element relative to earlier treatments is the systematic inclusion of \textit{collision-induced damping} arising from the imaginary parts of the GLFs. This allows us to analyse not only collective-mode dispersions, but also their intrinsic damping across generic angular-momentum channels at finite momenta ($\mathbf q$) and frequencies ($\mathbf \Omega$).
\item Relation to bosonization-based approaches:
\\Recent bosonization-based treatments of NFL states \cite{yrkv-42wd} emphasise on kinematic constraints and operator-level formulations of Fermi-surface fluctuations. While these methods provide valuable insights into universal scaling properties, they typically do not incorporate collision integrals or damping effects in a transparent kinetic-equation form. Our approach is complementary: it retains explicit momentum resolution along the Fermi surface and provides a direct kinetic description of collective modes, including damping, which is not readily accessible in current bosonization frameworks.
\item Relation to SYK-based models:
\\SYK-inspired models are based on large-$N$ formulation (e.g., Ref.~\cite{PhysRevB.103.235129}) and depend crucially on the large-$N$ limit. Their microscopic formulation is fundamentally different from working with the QFT-based action adopted here. We take the leading-order fermion-boson interaction describing the Ising-Nematic QCP and the controlled approximation dealing with this strongly-coupled theory is well-established within the dimensional-regularization scheme put forward in Refs.~\cite{Lee-Dalid, ips-uv-ir1}. Our QFT analysis has no dependence whatsoever on the number of flavours ($N$) of fermions and its validity is not contingent upon the large-$N$ limit. In particular, we have not found an exhaustive analysis of the collective modes of critical Fermi surfaces, including the full effects of the collision integrals like we have done, and neither is there a full explanation of the emergent collective modes quantified by their structure as well as $(\Omega, \, \mathbf q)$-dependence. In other words, our QBE framework is designed to address all these aspects in a holisitc way by incorporating the spatial structure, Fermi-surface geometry, energy-dependence, and finite-$\mathbf q $ dynamics, which are currently lacking in those SYK-based studies.
\item Novelty and limitations:
\\The principal new contributions of our work, beyond earlier works, can be summarised as
\begin{itemize}
\item inclusion of frequency-dependent \textit{complex} GLFs encoding collision-induced damping;
\item systematic analysis of collective modes in generic angular-momentum channels;
\item extension of the QBE framework to finite $(\Omega,\mathbf q)$ beyond the collisionless limit.
Limitations, if any, can be attributed to carrying out our analysis in the $T\rightarrow 0$ regime.
\end{itemize}
\end{enumerate} 
In the future, it will be worthwhile to extend our calculations to the finite-temperature regimes, which is a challenging task, given that we have to employ the framework of thermal quantum field theory. Other directions include recomputing everything for the case of a charged NFL, when Coulomb interactions will contribute (see, for example, Ref.~\cite{ips-zero-mode}).


\appendix
\section{Assorted Green’s functions in the Keldysh formalism}
\label{app1}

We study the collective modes of a critical Fermi surface at a quantum critical point using the Quantum Boltzmann equation (QBE). We use the standard definitions following Refs.~\cite{kadanoff2018quantum,mahan2013many,craig,rammer1986quantum} for the nonequilibrium matrix Green's functions. Denoting them as $\tilde{G}$ and $\tilde{D}$ for the fermions and the critical bosons, respectively, we have
\begin{align}
\tilde{G}=\begin{pmatrix}
	G_t & -G^<\\
	G^> & -G_{\bar{t}}
\end{pmatrix}, 
\quad
\tilde{D}=\begin{pmatrix}
	D_t & -D^<\\
	D^> & -D_{\bar{t}}
\end{pmatrix},
\label{eq-matrix-green}
\end{align}
where
\begin{align}
	G^>(x_1,x_2) & = -\,  i \, \langle \psi(x_1)\, \psi^\dagger(x_2) \rangle \,, \quad
	G^<(x_1,x_2) = i \, \langle \psi^\dagger(x_2)\, \psi(x_1) \rangle \,, \nn
	G_t(x_1,x_2) & = \Theta(t_1-t_2) \, G^>(x_1,x_2)  +  \Theta(t_2-t_1) \, G^<(x_1,x_2) \,,\nn
	G_{\bar{t}}(x_1,x_2) & =\Theta(t_2-t_1) \, G^>(x_1,x_2)  +  \Theta(t_1-t_2) \, G^<(x_1,x_2) \,,
\end{align}
and 
\begin{align}
	D^>(x_1,x_2) & = - \, i \, \langle \phi(x_1)\, \phi({x_2}) \rangle\,,\quad
	D^<(x_1,x_2) = - \, i \, \langle \phi(x_2)\, \phi({x_1}) \rangle, \\
	D_t(x_1,x_2) & = \Theta(t_1-t_2) \, D^>(x_1, x_2)  +
	\Theta(t_2-t_1) \, D^<(x_1,x_2) \, ,\\
	D_{\bar{t}}(x_1,x_2) & = \Theta(t_2-t_1) \, D^>(x_1,x_2)  +  \Theta(t_1-t_2) \, D^<(x_1,x_2) \,.
\end{align}
Here, $x_1 \equiv (t_1, {\mathbf r}_1)$ and $x_2 \equiv (t_2, {\mathbf r}_2)$ (using the condensed notation for the spacetime coordinates), and $\psi (x)$ [$\phi(x)$] is the the fermionic [bosonic] quantum field. The Green's functions shown above are related to the retarded ($G^R$), advanced ($G^A$), and Keldysh ($G^K$) Green's functions in the following way:
\begin{align}
	G^R & = G_t -G^< =G^> -G_{\bar{t}} \,, \quad
	G^A =G_t -G^> =G^< -G_{\bar{t}} \,, \quad
	G^K =G^<  + G^> \,,
	\label{eq-gegkgless}
\end{align}
where, for notational convenience, we have dropped the functional dependence on $x_1$ and $ x_2$. A straightforward manipulation, using Eq.~\eqref{eq-gegkgless}, leads to
\begin{align}
	G_t & = \frac{G^< + G^>}{2}  +  \text{Re} [G^R ]\,, \quad
	G_{\hat{t}} =\frac{G^< + G^>}{2}- \text{Re} [G^R] \,.
\end{align}
Analogous relations hold for the critical bosons as well. The fermionic and bosonic self-energy matrices, $\tilde{\Sigma}$ and $\tilde{\Pi}$, have a $2\times2$ matrix structure, analogous to Eq.~\eqref{eq-matrix-green}. They follow the Dyson equations, viz.
\begin{align}
	(G_0^{-1}-\tilde{\Sigma}) \circ  \tilde{G} & = 1 \,, \quad
	(D_0^{-1}-\tilde{\Pi}) \circ  \tilde{D} =1.
	\label{eq-fer-bos}
\end{align}
Here, the symbol $ \circ $ represents multiplication both in spacetime and matrix structure [for example, we have $A \circ  B = \int dx_2 \, A_{i,j}(x_1,x_2) \, B_{j,k}(x_2,x_3)$]. $G_0$ and $D_0$ are the noninteracting Green's functions for the fermions and bosons, respectively, which evaluate to
\begin{align}
	G_0^{-1}(\omega_k , \bk ) & = \omega_k - \xi_\bk  \text{ and }
	D_0^{-1}(\omega_k , \bk ) =\omega_k ^2-\bk^2 .
\end{align}
Here, $\xi_\bk$ denotes the bare fermionic dispersion. 

\section{Equilibrium Green's functions and spectral properties} 
\label{appeq}

For a generic system of fermions and bosons in equilibrium, the Fourier-space Green's functions can be expressed as follows:
\begin{align}
G_0^<(\omega_k, \bk ) & = i \, f_0(\omega_k) \,A (\omega_k, \bk ) \,,\quad
G_0^>(\omega_k, \bk )  =  -\, i \, [ 1- f_0(\omega_k)]\,  A(\omega_k, \bk )\,,\nn
D_0^<(\omega_k, \bk ) & = -\, i \, \, n_B(\omega_k) \, B(\omega_k, \bk )\,, \quad
D_0^>(\omega_k, \bk )  =  -\, i \, [ 1 + n_B(\omega_k) ]\,B(\omega_k, \bk )\,,
\label{eq-D-more-less}
\end{align}
where $ f_0(\omega) = ( e^{\beta \, \omega} + 1 )^{-1} $ and $n_B (\omega) = ( e^{\beta \, \omega} - 1 )^{-1}  $ are the Fermi-Dirac and Bose-Einstein distributions functions, respectively, at temperature $T=1/\beta$. While $A(\omega_k, \bk )$ and $B(\omega_k, \bk ) $ denote the spectral functions for the fermions and bosons, respectively, we have used a subscript ``0'' to denote the equilibrium state. 

Applying Eq.~\eqref{eq-fer-bos} to the equilibrium condition, the standard expressions for the retarded Green's functions turn out to be
\begin{align}
G_0^R(\omega_k, \bk )   =  \frac{1} {\omega_k -\xi_\bk-\Sigma_0^R(\omega_k, \bk )} \text{ and }
D_0^R(\omega_k, \bk )   =  \frac{1} {\omega_k^2-|\bk|^2-\Pi_0^R(\omega_k, \bk )}\,,
\label{eq-bos-propr}
\end{align} 
which include the effects of the retarded self-energies, $\Sigma_0^R$ and $\Pi_0^R$.
The spectral functions are related to these Green's functions through $A(\omega_k, \bk )
= - \,2 \,\text{Im}[G_0^R(\omega_k, \bk ) ]$ and  $ B(\omega_k, \bk )=- \,2 \,
\text{Im} [D_0^R(\omega_k, \bk )]$, leading to
\begin{align}
	\label{eqspec}
	A(\omega_k, \bk ) & = 
	\frac{ - \,2  \,  \text{Im} [\Sigma_0^R(\omega_k, \bk )] }
	{\left[\omega_k - \xi_\bk-\text{Re}[ \Sigma_0^R(\omega_k, \bk )] \right]^2 
		+  \left[\text{Im}\Sigma_0^R(\omega_k, \bk )\right]^2} \,, \nn
	B(\omega_k, \bk ) & = 
	\frac{- \,2 \, \text{Im} [\Pi_0^R (\omega_k, \bk )] }
	{\left[\omega_k^2-|\bk|^2 -\text{Re} [\Pi^R (\omega_k, \bk )] \right]^2 
		+  \left[\text{Im} [ \Pi_0^R(\omega_k, \bk )] \right]^2} \,.
\end{align}

\section{Fermionic part of the QBEs}
\label{appfer}

In order to derive the QBE, we employ the Dyson equation for the fermions [cf. Eq.~\eqref{eq-fer-bos}], relevant for the $G^<$ component. While the details can be found in Refs.~\cite{kim_qbe, prange, ips-kazi, ips-zero-mode}, we present a brief outline for the convenience of the reader and, also, for setting up the notations. We multiply both the sides of Eq.~\eqref{eq-fer-bos} with $\tilde{G}$ from the left and $\tilde{G}^{-1}$ from the right, subtract the result from Eq.~\eqref{eq-fer-bos}, and write the resulting $(1,2)^{\rm th}$ matrix-element as  
\begin{align}
	G_0^{-1}\circ G^<-G^<\circ G_0^{-1} & = 
	\text{Re}[\Sigma^R] \circ G^<-G^<\circ \text{Re}[\Sigma^R]  
	+  \Sigma^<\circ \text{Re}[G^R]-\text{Re} [G^R] \circ \Sigma^<
	\nn & \;\quad 
	+  \frac{  \Sigma^>\circ G^< 
		+ G^<\circ \Sigma^>-\Sigma^<\circ G^>-G^>\circ \Sigma^< } {2}\,.
	\label{eq-G-less}
\end{align}
Here, the equilibrium Green's functions and self-energies have been indicated by using the subscript ``$0$''. Next, we perform Fourier transformations on Eq.~\eqref{eq-G-less}, from the center-of-mass coordinates coordinates to their canonically-conjugate variables (see the discussions in Sec.~\ref{seckeldysh}). This leads to
\begin{align}
	&   \left[G_0^{-1}(k)-\text{Re}[\Sigma^R ( k, x_{\rm rel}) ], G^< ( k, x_{\rm rel}) \right]_{\text{PB}}
	- \left[\Sigma^< ( k, x_{\rm rel}) ,\text{Re}[G^R ( k, x_{\rm rel}) ] \right]_{\text{PB}}
	\nn &    =
	-i \left[ \Sigma^> ( k, x_{\rm rel})  \,G^< ( k, x_{\rm rel})   
	- \Sigma^< ( k, x_{\rm rel})  \, G^> ( k, x_{\rm rel}) \right ], 
	\label{{eq-G-wt}}
\end{align}
where \begin{align}
	\left[A ( k, x_{\rm rel}) , B ( k, x_{\rm rel}) \right]_{\text{PB}} & \equiv
	\sum_{i=1}^3 \left [   \partial_{x_i} A\, \partial_{k_i} B 
	- \partial_{ k_i} A\, \partial_{x_i} B \right ]
	= \nabla_{\mathbf r } A\cdot \nabla_{\mathbf p }  B
	- \nabla_{\mathbf k } A \cdot \nabla_{\mathbf r }  B
	+ \partial_\omega A \, \partial_t B
	- \partial_t A \, \partial_\omega B \,.
	\label{eqPB}
\end{align}

In order to study the dynamics not too far from the equilibrium, we linearise the resulting equation in the deviations parametrised by $\delta G^< \equiv G^<-G^<_0$, $\delta \text{Re}[G^R] \equiv \text{Re}[G^R] - \text{Re}[G^R_0]$, $\delta \Sigma^{<(>)}\equiv \Sigma^{<(>)}-\Sigma^{<(>)}_0$, and $\delta \text{Re}[\Sigma^R ] \equiv \text{Re}[\Sigma^R] 
- \text{Re}[\Sigma^R_0]$. All these exercises result in
\begin{align}
	\label{QBEf}  
	&  \left[G_0^{-1}( p  +   q/2)-G_0^{-1}( p- q/2)\right] \delta G^<(p,q)
	-\Big [ \text{Re} [ \Sigma_0^R( p  +  q/2) ]  -\text{Re}  [\Sigma_0^R(p-q/2)]
	\Big ] \, \delta G^<(p,q)  
	\nn  &
	+  \, \left[G_0^<( p  +  q/2)-G_0^<( p- q/2)\right]
	\delta \Big (\text{Re} [ \Sigma^R(p,q)]\Big )
	-\left[\Sigma_0^<( p + q/2) - \Sigma_0^<( p-q/2)\right]\delta \Big (\text{Re} [ G^R(p,q)] \Big ) 
	\nn & 
	+    \, \Big [\text{Re} [ G_0^R( p + q/2)]-\text{Re} [ G_0^R( p-q/2)]
	\Big] \, \delta \Sigma^<(p,q) 
	= I_{\text{coll}} (p,q)\,,\nn &
	\nn & \text{where }
	I_{\text{coll}} (p, q) = 
	G_0^<(p)\, \delta \Sigma^>(p,q)  +  \Sigma_0^>(p)  \,\delta G^<(p,q)
	-G_0^>(p)  \,  \delta \Sigma^<(p,q)-\Sigma_0^<(p)   \,   \delta G^>(p,q)  \,.
\end{align}
Here, $p$ and $q$ are connected with the relative and the center-of-mass spacetime coordinates, respectively. 
Integrating both sides with respect to $\xi_\bp$, so that we can use the definitions in Eq.~\eqref{eq:gdf2}, we arrive at
\begin{align}
	& (\Omega-v_F \,|\bq|\cos\theta_{\bp\bq}) \, \delta f ( \omega_p , \theta_{\bp \bq}; \Omega, \bq)
	-\left[\text{Re}[\Sigma_0^R \Big (\omega_p  +  \frac{\Omega}{2} \Big)]
	-\text{Re} [\Sigma_0^R \Big(\omega_p-\frac{\Omega}{2} \Big)] \right] 
	\delta f ( \omega_p , \theta_{\bp \bq}; \Omega, \bq)
	\nn &  \hspace{ 5.5 cm} +  \left[f_0 \Big (\omega_p  +  \frac{\Omega}{2} \Big )
	- f_0 \Big (\omega_p-\frac{\Omega}{2} \Big )\right] 
	\delta \text{Re} [ \Sigma^R ( \omega_p, \theta_{\bp \bq}; \Omega, \bq)]
	= \int \frac{d\xi_\bp}{ 2 \, \pi } 
	\,  I_{\text{coll}} ( \omega_p, \theta_{\bp \bq}; \Omega, \bq)\,,
	\label{eq_int_qbe}
\end{align}
where $\theta_{\bp \bq}=\theta_\bp-\theta_\bq$, and
\begin{align}
	\int \frac{d\xi_\bp}{ 2 \, \pi } \,
	I_{\text{coll}} ( \omega_p, \theta_{\bp \bq}; \Omega, \bq) 
	&   = f_0(\omega_p) \,\delta\Sigma^> ( \omega_p, \theta_{\bp \bq}; \Omega, \bq)  
	+  \Sigma_0^>(\omega_p) \, \delta f ( \omega_p, \theta_{\bp \bq}; \Omega, \bq) 
	\nn & \,\quad
	-  [ f_0(\omega_p)-1 ] \, \delta \Sigma^< ( \omega_p, \theta_{\bp \bq}; \Omega, \bq) 
	-\Sigma_0^<(\omega_p) \, \delta f ( \omega_p, \theta_{\bp \bq}; \Omega, \bq) \,.
	\label{eq-int-coll}
\end{align} 

In order to progress further, we need the expressions for the self-energies. Starting with
$$
\Sigma^<(x_1,x_2)  = i\, e^2\, G^<(x_1,x_2) \, D^>(x_1,x_2)\,, $$
the Fourier-transformed counterpart takes the form of
\begin{align}
	\Sigma^< ( \omega_p, \theta_{\bp \bq}; \Omega, \bq)  & = 
	i\, e^2 \int_{-\infty}^{\infty} \frac{d\nu} { 2 \, \pi }\,
	\int\frac{d^2 \bk}   {( 2 \, \pi )^2} \, 
	G^<( \omega_p +\nu, \bp + \bk  ;\Omega, \bq) 
	\, D^>(\nu ,\bk,;\Omega, \bq)
	\nn & = i \,e^2\, N_0 \int_{0}^{\infty} \frac{d\nu}{ 2 \, \pi }
	\int \frac{d\xi_{\bk}}{ 2 \, \pi } 
	\int \frac{d\theta_{\bk \bq}}  { 2 \, \pi } \,\Big[
	G^<( \omega_p +\nu  ,\xi_{\bk}, \theta_{\bk \bq} ; \Omega, \bq) \, 
	D^>( \nu , \bk-\bp ;  \Omega, \bq )
	\nn & \hspace{ 5.35 cm }
	+ G^<( \omega_p  - \nu  ,\xi_{\bk}, \theta_{\bk \bq} ; \Omega, \bq ) \, 
	D^<( \nu , \bk-\bp;  \Omega, \bq ) \Big ]\,,
	\label{sigma-eq2}
\end{align}
where $N_0$ is the fermionic density-of-states at the Fermi surface. Also, we have used the property, $D^>( k, x_{\rm rel}) =
D^<(- k, x_{\rm rel})$, applicable for real bosons.

\section{Fermionic part of the QBEs with the bosons assumed to be in equilibrium}
\label{appfer0}

In this appendix, we show the form of the fermionic QBE [viz. Eq.~\eqref{eq_int_qbe}] when the bosons are assumed to be in equilibrium. In this situation, the bosonic propagators do not depend on $\mathbf q$, and the relations shown in Eq.~\eqref{eq-D-more-less} are applicable. After carrying out the $\xi_\bk$-integration, using Eq.~\eqref{eq:gdf2}, Eq.~\eqref{sigma-eq2} simplifies to
\begin{align}
\frac{ \Sigma^< ( \omega_p, \theta_{\bp \bq}; \Omega, \bq) }
{ -\, i \, \, N_0\, e^2 } =  
\int \frac{d\theta_{\bk \bq}} {2\,\pi}
\int_0^\infty \frac{d\nu}{\pi} \,
\text{Im} [D_0^R( \nu, \bk-\bp )] \left[n_B(\nu) \,f( \omega_p -\nu, \theta_{\bk \bq} ;  \Omega, \bq ) 
+  \lbrace n_b(\nu) + 1 \rbrace 
f( \omega_p   +  \nu, \theta_{\bk \bq};   \Omega, \bq ) \right].
\label{sigma less in equilibrium}
\end{align}
A similar calculation, using $ \Sigma^>(x_1,x_2) = i \,e^2 \, G^>(x_1,x_2) \, D^<(x_1,x_2) $, yields
\begin{align}
& \frac{ \Sigma^>(  \omega_p, \theta_{\bp\bq} ;  \Omega, \bq ) }
{- \, i \, \, N_0\, e^2}
\nn & =    
\int \frac{d\theta_{\bk \bq}}{2\,\pi}
\int_0^\infty \frac{d\nu}{\pi} \, \text{Im} [D_0^R( \nu , \bk-\bp)] 
\left[ n_B(\nu) \left \lbrace f( \omega_p +\nu, \theta_{\bk \bq};  \Omega, \bq )-1 \right \rbrace
+  \lbrace n_B (\nu)  +  1 \rbrace  
\left \lbrace 
f( \omega_p -\nu, \theta_{\bk \bq};   \Omega, \bq )-1\right \rbrace  \right].
\label{sigma great in equilibrium}
\end{align}
Using the Kramers–Kronig relations, we calculate the retarded self-energy to be
\begin{align}
\text{Re} [\Sigma^R ( \omega_p, \theta_{\bp \bq}; \Omega, \bq) ]
& \equiv 
-\int \frac{d\omega'}{\pi} \,
P \, \frac{\text{Im}[\Sigma^R ( \omega', \theta_{\bp\bq};  \Omega, \bq )]} 
{\omega_p-\omega'}
= \int \frac{d\omega'}{ 2 \, \pi \, i} 
\, P  \frac{\Sigma^< (\omega' ,\theta_{\bp\bq};  \Omega, \bq )
	-\Sigma^> ( \omega' , \bp;  \Omega_p, \bq )}  {\omega_p-\omega'}
\nn &   = - \, N_0\, e^2  
\int \frac{d\theta_{\bk \bq}} {2\,\pi}
\int_{-\infty}^\infty \frac{d\nu}{ 2 \, \pi } \,
\text{Re} [ D_0^R( \nu-\omega_p , \bk-\bp) ] 
\,f( \nu, \theta_{\bk \bq} ;  \Omega, \bq )\,,
\label{eqre-self-en-eqm}
\end{align}
where we have used Eqs.~\eqref{sigma less in equilibrium} and \eqref{sigma great in equilibrium}, and the symbol $P$ denotes the principal value. 

In an attempt to further simplify, we note that $D_0^R(\nu, \bk-\bp)$ [cf. Eq.~\eqref{listgfbos}] depends on the magnitude of the exchange momentum, $|\bk-\bp| \simeq  2 \, k_F |\sin (\theta_{\bk\bp}/2)|$. Thus, it can approximated by
\begin{align}
	& D_{0}^{R}(\nu,\bk-\bp) 
	\simeq D_{0}^{R}(\nu, e_{\bk \bp}) 
	= -\, \frac{1}{ e_{\bk \bp}^2- i \, \, \chi \, {\nu} / e_{\bk \bp}}\,,
	\quad
	e_{\bk \bp} \equiv  2 \, k_F \,|\sin ( {\theta_{\bk\bp}}/ 2 )|\,,\nn
	& \Rightarrow
	\text{Re} [D_{0}^R(\nu, e_{\bk \bp})] =
	- \, \frac{e_{\bk \bp}^4} {e_{\bk \bp}^6  +  \chi^2 \,\nu^2} \,,\quad 
	\text{Im} [D_{0}^R(\nu,  e_{\bk \bp})]
	= -\, \frac{ \chi  \, e_{\bk \bp}\, \nu} 
	{e_{\bk \bp}^6  +  \chi^2 \,\nu^2}\,.
\label{reimD}
\end{align}

Now, we are in a position to plug the expressions for the self-energies into the left-hand side of Eq.~\eqref{eq_int_qbe}, which reduces to
\begin{align}
	\label{eqlhs}  
	& (\Omega-v_F |\bq| \cos\theta_{\bp \bq}) \, \delta f ( \omega_p, \theta_{\bp \bq}; \Omega, \bq) 
	\nn & 
	+ e^2 \, N_0 \int \frac{ d\theta_{\bp'\bq}}{ 2 \, \pi }
	\int_{-\infty}^{\infty} \frac{d\omega'} { 2 \, \pi } \,
	\text{Re} [ D^R_0( \omega'-\omega_p, \bp-\bp') ] 
	\left[ f_0 \Big (\omega'  +  \frac{\Omega}{2} \Big ) 
	- f_0 \Big (\omega'-\frac{\Omega}{2} \Big) \right]
	\delta f ( \omega_p , \theta_{\bp \bq}; \Omega, \bq) 
	\nn &  
	- e^2 \, N_0 \int \frac{d\theta_{\bp'\bq}}{ 2 \, \pi }
	\int_{-\infty}^{\infty} \frac{d\omega'}{ 2 \, \pi } \,
	\text{Re} [D^R_0( \omega'-\omega_p, \bp-\bp') ]
	\left[ f_0 \Big (\omega_p  +  \frac{\Omega}{2} \Big) -f_0 \Big (\omega_p - \frac{\Omega}{2} \Big )\right] 
	\delta f( \omega', \theta_{\bp'\bq} ;  \Omega, \bq )\,. 
\end{align}
Similarly, the right-hand side [viz. Eq.~\eqref{eq-int-coll}] reduces to
\begin{align}
	\label{eqrhs}   
	i \, e^2 \, N_0 \int \frac{d\theta_{\bp'\bq}} { 2 \, \pi } \int_{0}^{\infty}
	\frac{d\nu}{\pi}\, \text{Im} [D_0^R( \nu , \bp-\bp')]
	\int_{-\infty}^{\infty} d\omega' 
	\Big [ &
	\delta(\omega'-\omega_p +  \nu) \,
	\delta f( \omega', \theta_{\bp'\bq};  \Omega, \bq ) 
	\left \lbrace n_B(\nu)  + f_0(\omega_p) \right \rbrace
	\nn & 
	- \delta f ( \omega_p , \theta_{\bp \bq}; \Omega, \bq)
	\left \lbrace 1 + n_B(\nu)-f_0(\omega') \right \rbrace
	\nn &
	+  \delta(\omega'-\omega_p - \nu) 
	\, \delta f( \omega', \theta_{\bp'\bq};  \Omega, \bq )
	\left \lbrace  1 + n_B(\nu)-f_0(\omega_p) \right \rbrace
	\nn &   
	-  \delta(\omega'-\omega_p - \nu) \, 
	\delta f ( \omega_p, \theta_{\bp \bq}; \Omega, \bq)
	\left \lbrace  n_B(\nu) + f_0(\omega') \right \rbrace  \Big ]\,.
\end{align}

\bibliography{bibqbe}

@article{khveshchenko,
  title = "{Collective modes in two-dimensional non-Fermi liquids}",
  author = {Khveshchenko, D. V.},
  journal = {Phys. Rev. B},
  volume = {111},
  issue = {16},
  pages = {L161108},
  numpages = {5},
  year = {2025},
  month = {Apr},
  publisher = {American Physical Society},
  doi = {10.1103/PhysRevB.111.L161108},
  url = {https://link.aps.org/doi/10.1103/PhysRevUltravioletB.111.L161108}
}

@article{PhysRevB.103.235129,
  title = {Large-{$N$} theory of critical {F}ermi surfaces},
  author = {Esterlis, Ilya and Guo, Haoyu and Patel, Aavishkar A. and Sachdev, Subir},
  journal = {Phys. Rev. B},
  volume = {103},
  issue = {23},
  pages = {235129},
  numpages = {32},
  year = {2021},
  month = {Jun},
  publisher = {American Physical Society},
  doi = {10.1103/PhysRevB.103.235129},
  url = {https://link.aps.org/doi/10.1103/PhysRevB.103.235129}
}

@article{yrkv-42wd,
  title = {Fermi surface bosonization for non-{F}ermi liquids},
  author = {Han, SangEun and Desrochers, F\'elix and Kim, Yong Baek},
  journal = {Phys. Rev. B},
  volume = {112},
  issue = {16},
  pages = {165116},
  numpages = {25},
  year = {2025},
  month = {Oct},
  publisher = {American Physical Society},
  doi = {10.1103/yrkv-42wd},
  url = {https://link.aps.org/doi/10.1103/yrkv-42wd}
}

@article{metzner,
  title = {Fluctuation effects at the onset of the $2{k}_{F}$ density wave order with one pair of hot spots in two-dimensional metals},
  author = {S\'ykora, J\'achym and Holder, Tobias and Metzner, Walter},
  journal = {Phys. Rev. B},
  volume = {97},
  issue = {15},
  pages = {155159},
  numpages = {13},
  year = {2018},
  month = {Apr},
  publisher = {American Physical Society},
  doi = {10.1103/PhysRevB.97.155159},
  url = {https://link.aps.org/doi/10.1103/PhysRevB.97.155159}
}

@article{ips-2kf,
    author = "Mandal, Ipsita",
    title = "{Stable non-Fermi liquid fixed point at the onset of incommensurate 2$k_F$ charge density wave order}",
    doi = "10.1016/j.nuclphysb.2024.116586",
    journal = "Nucl. Phys. B",
    volume = "1005",
    pages = "116586",
    year = "2024"
}

@ARTICLE{ips-hermann-review,
title = "{Transport properties in non-Fermi liquid phases of nodal-point semimetals}",
author = {Ipsita Mandal and Hermann Freire},
doi = {10.1088/1361-648X/ad665e},
url = {https://dx.doi.org/10.1088/1361-648X/ad665e},
year = {2024},
month = {aug},
publisher = {IOP Publishing},
volume = {36},
number = {44},
pages = {443002},
journal = {Journal of Physics: Condensed Matter}
}

@article{yoichi,
  title = {Electrical Resistivity Anisotropy from Self-Organized One Dimensionality in High-Temperature Superconductors},
  author = {Ando, Yoichi and Segawa, Kouji and Komiya, Seiki and Lavrov, A. N.},
  journal = {Phys. Rev. Lett.},
  volume = {88},
  issue = {13},
  pages = {137005},
  numpages = {4},
  year = {2002},
  month = {Mar},
  publisher = {American Physical Society},
  doi = {10.1103/PhysRevLett.88.137005},
  url = {https://link.aps.org/doi/10.1103/PhysRevLett.88.137005}
}

@article{Kohsaka,
author = {Y. Kohsaka  and C. Taylor  and K. Fujita  and A. Schmidt  and C. Lupien  and T. Hanaguri  and M. Azuma  and M. Takano  and H. Eisaki  and H. Takagi  and S. Uchida  and J. C. Davis },
title = {An Intrinsic Bond-Centered Electronic Glass with Unidirectional Domains in Underdoped Cuprates},
journal = {Science},
volume = {315},
number = {5817},
pages = {1380-1385},
year = {2007},
doi = {10.1126/science.1138584},
URL = {https://www.science.org/doi/abs/10.1126/science.1138584}
}

@article{debanjan,
  title = {Collective density fluctuations of strange metals with critical {F}ermi surfaces},
  author = {Wang, Xuepeng and Chowdhury, Debanjan},
  journal = {Phys. Rev. B},
  volume = {107},
  issue = {12},
  pages = {125157},
  numpages = {17},
  year = {2023},
  month = {Mar},
  publisher = {American Physical Society},
  doi = {10.1103/PhysRevB.107.125157},
  url = {https://link.aps.org/doi/10.1103/PhysRevB.107.125157}
}

@article{else,
  title = {Collisionless dynamics of general non-{F}ermi liquids from hydrodynamics of emergent conserved quantities},
  author = {Else, Dominic V.},
  journal = {Phys. Rev. B},
  volume = {108},
  issue = {4},
  pages = {045107},
  numpages = {13},
  year = {2023},
  month = {Jul},
  publisher = {American Physical Society},
  doi = {10.1103/PhysRevB.108.045107},
  url = {https://link.aps.org/doi/10.1103/PhysRevB.108.045107}
}

@book{mahan2013many,
  title={Many-Particle Physics},
  author={Mahan, Gerald D.},
  isbn={9781475757149},
  series={Physics of Solids and Liquids},
  url={https://books.google.co.in/books?id=TFDUBwAAQBAJ},
  year={2013},
  publisher={Springer US}
}

@book{sachdev_2011, 
place={Cambridge}, edition={2}, title={Quantum Phase Transitions},
DOI={10.1017/CBO9780511973765}, publisher={Cambridge University Press}, 
author={Sachdev, Subir}, year={2011}
}

@article{hinkov,
author = {V. Hinkov  and D. Haug  and B. Fauqué  and P. Bourges  and Y. Sidis  and A. Ivanov  and C. Bernhard  and C. T. Lin  and B. Keimer },
title = {Electronic Liquid Crystal State in the High-Temperature Superconductor \ce{YBa2Cu3O}$_{6.45}$},
journal = {Science},
volume = {319},
number = {5863},
pages = {597-600},
year = {2008},
doi = {10.1126/science.1152309},
URL = {https://www.science.org/doi/abs/10.1126/science.1152309}
}

@article{olav,
  title = {Transverse Gauge Interactions and the Vanquished {F}ermi Liquid},
  author = {Chakravarty, Sudip and Norton, Richard E. and Sylju\aa{}sen, Olav F.},
  journal = {Phys. Rev. Lett.},
  volume = {74},
  issue = {8},
  pages = {1423--1426},
  numpages = {0},
  year = {1995},
  month = {Feb},
  publisher = {American Physical Society},
  doi = {10.1103/PhysRevLett.74.1423},
  url = {https://link.aps.org/doi/10.1103/PhysRevLett.74.1423}
}

@article{Daou,
   title={Broken rotational symmetry in the pseudogap phase of a high-{T}$_c$ superconductor},
   volume={463},
   ISSN={1476-4687},
   url={http://dx.doi.org/10.1038/nature08716},
   DOI={10.1038/nature08716},
   number={7280},
   journal={Nature},
   publisher={Springer Science and Business Media LLC},
   author={Daou, R. and Chang, J. and LeBoeuf, David and Cyr-Choinière, Olivier and Laliberté, Francis and Doiron-Leyraud, Nicolas and Ramshaw, B. J. and Liang, Ruixing and Bonn, D. A. and Hardy, W. N. and et al.},
   year={2010},
   month={Jan},
   pages={519–522}
}

@article{prange,
  title = {Transport Theory for Electron-Phonon Interactions in Metals},
  author = {Prange, Richard E. and Kadanoff, Leo P.},
  journal = {Phys. Rev.},
  volume = {134},
  issue = {3A},
  pages = {A566--A580},
  numpages = {0},
  year = {1964},
  month = {May},
  publisher = {American Physical Society},
  doi = {10.1103/PhysRev.134.A566},
  url = {https://link.aps.org/doi/10.1103/PhysRev.134.A566}
}

@Article{ips-uv-ir2,
   author = "Mandal, Ipsita",
    title = "{UV/IR mixing in non-Fermi liquids: Higher-loop corrections in different energy ranges}",
    doi = "10.1140/epjb/e2016-70509-4",
    journal = "Eur. Phys. J. B",
    volume = "89",
    number = "12",
    pages = "278",
    year = {2016}
}

@Article{ips-uv-ir1,
  title = {Ultraviolet/infrared mixing in non-{F}ermi liquids},
  author = {Mandal, Ipsita and Lee, Sung-Sik},
  journal = {Phys. Rev. B},
  volume = {92},
  issue = {3},
  pages = {035141},
  numpages = {15},
  year = {2015},
  month = {Jul},
  publisher = {American Physical Society},
  doi = {10.1103/PhysRevB.92.035141},
  url = {https://link.aps.org/doi/10.1103/PhysRevB.92.035141}
}

@article{kim_qbe,
  title = {Quantum {B}oltzmann equation of composite fermions interacting with a gauge field},
  author = {Kim, Yong Baek and Lee, Patrick A. and Wen, Xiao-Gang},
  journal = {Phys. Rev. B},
  volume = {52},
  issue = {24},
  pages = {17275--17292},
  numpages = {0},
  year = {1995},
  month = {Dec},
  publisher = {American Physical Society},
  doi = {10.1103/PhysRevB.52.17275},
  url = {https://link.aps.org/doi/10.1103/PhysRevB.52.17275}
}

@book{kamenev, 
place={Cambridge}, 
title={Field Theory of Non-Equilibrium Systems}, 
DOI={10.1017/CBO9781139003667}, 
publisher={Cambridge University Press}, 
author={Kamenev, Alex}, 
year={2011}
}

@article{ips-rafael,
  title = {Valley-polarized nematic order in twisted moir\'e systems: In-plane orbital magnetism and crossover from non-{F}ermi liquid to {F}ermi liquid},
  author = {Mandal, Ipsita and Fernandes, Rafael M.},
  journal = {Phys. Rev. B},
  volume = {107},
  issue = {12},
  pages = {125142},
  numpages = {13},
  year = {2023},
  month = {Mar},
  publisher = {American Physical Society},
  doi = {10.1103/PhysRevB.107.125142},
  url = {https://link.aps.org/doi/10.1103/PhysRevB.107.125142}
}

@article{ips-hermann3,
doi = {10.1088/1361-648x/ac6785},
	url = {https://doi.org/10.1088/1361-648x/ac6785},
	year = 2022,
	month = {may},
	publisher = {{IOP} Publishing},
	volume = {34},
	number = {27},
	pages = {275604},
	author = {Ipsita Mandal and Hermann Freire},
	title = {Raman response and shear viscosity in the non-{F}ermi liquid phase of {L}uttinger semimetals},
	journal = {Journal of Physics: Condensed Matter}
}

@article{ips-hermann2,
title = {Thermoelectric and thermal properties of the weakly disordered non-{F}ermi liquid phase of {L}uttinger semimetals},
journal = {Physics Letters A},
volume = {407},
pages = {127470},
year = {2021},
issn = {0375-9601},
doi = {https://doi.org/10.1016/j.physleta.2021.127470},
url = {https://www.sciencedirect.com/science/article/pii/S0375960121003340},
author = {Hermann Freire and Ipsita Mandal}
}

@ARTICLE{ips-hermann,
  title = {Transport in the non-{F}ermi liquid phase of isotropic {L}uttinger semimetals},
  author = {Mandal, Ipsita and Freire, Hermann},
  journal = {Phys. Rev. B},
  volume = {103},
  issue = {19},
  pages = {195116},
  numpages = {12},
  year = {2021},
  month = {May},
  publisher = {American Physical Society},
  doi = {10.1103/PhysRevB.103.195116},
  url = {https://link.aps.org/doi/10.1103/PhysRevB.103.195116}
}

@article{Abrikosov,
  title = {Calculation of critical indices for zero-gap semiconductors},
  author = {Abrikosov, A. A.},
  journal = {Sov. Phys.-JETP},
  volume = {39},
  pages = {709},
  numpages = {8},
  year = {1974},
  url={https://www.jetp.ras.ru/cgi-bin/e/index/e/39/4/p709?a=list}
}

@article{malcolm-bitan,
  title = {From Birefringent Electrons to a Marginal or Non-{F}ermi Liquid of Relativistic Spin-$1/2$ fermions: {A}n Emergent Superuniversality},
  author = {Roy, Bitan and Kennett, Malcolm P. and Yang, Kun and Juri\ifmmode \check{c}\else \v{c}\fi{}i\ifmmode \acute{c}\else \'{c}\fi{}, Vladimir},
  journal = {Phys. Rev. Lett.},
  volume = {121},
  issue = {15},
  pages = {157602},
  numpages = {5},
  year = {2018},
  month = {Oct},
  publisher = {American Physical Society},
  doi = {10.1103/PhysRevLett.121.157602},
  url = {https://link.aps.org/doi/10.1103/PhysRevLett.121.157602}
}

@article{rahul-sid,
  title = {Disorder-driven destruction of a non-{F}ermi liquid semimetal studied by renormalization group analysis},
  author = {Nandkishore, Rahul M. and Parameswaran, S. A.},
  journal = {Phys. Rev. B},
  volume = {95},
  issue = {20},
  pages = {205106},
  numpages = {14},
  year = {2017},
  month = {May},
  publisher = {American Physical Society},
  doi = {10.1103/PhysRevB.95.205106},
  url = {https://link.aps.org/doi/10.1103/PhysRevB.95.205106}
}

@ARTICLE{polchinski,
   author = {{Polchinski}, J.},
    title = "{Low-energy dynamics of the spinon-gauge system}",
  journal = {Nuclear Physics B},
     year = 1994,
    month = jul,
   volume = 422,
    pages = {617-633},
      doi = {10.1016/0550-3213(94)90449-9},
   adsurl = {http://adsabs.harvard.edu/abs/1994NuPhB.422..617P}
}

@article{ALTSHULER,
  title = {Low-energy properties of fermions with singular interactions},
  author = {Altshuler, B. L. and Ioffe, L. B. and Millis, A. J.},
  journal = {Phys. Rev. B},
  volume = {50},
  issue = {19},
  pages = {14048--14064},
  numpages = {0},
  year = {1994},
  month = {Nov},
  publisher = {American Physical Society},
  doi = {10.1103/PhysRevB.50.14048},
  url = {http://link.aps.org/doi/10.1103/PhysRevB.50.14048}
}

@article{ips-birefringent,
      title = {Robust marginal {F}ermi liquid in birefringent semimetals},
journal = {Physics Letters A},
volume = {418},
pages = {127707},
year = {2021},
issn = {0375-9601},
doi = {https://doi.org/10.1016/j.physleta.2021.127707},
url = {https://www.sciencedirect.com/science/article/pii/S0375960121005715},
author = {Ipsita Mandal}
}

@article{ips-cavity,
title = {Induction of non-{F}ermi liquids by critical cavity photons at the onset of superradiance},
author = {Ipsita Mandal},
journal = {Annals of Physics},
volume = {474},
pages = {169925},
year = {2025},
issn = {0003-4916},
doi = {https://doi.org/10.1016/j.aop.2025.169925},
url = {https://www.sciencedirect.com/science/article/pii/S0003491625000065}
}

@article{ips-cdw-moire,
       author = {{Mandal}, Ipsita},
        title = "{Non-Fermi liquid behaviour of CDW instabilities in fractionally-filled moir{\'e} flatbands}",
   journal = {Annals of Physics},
volume = {486},
pages = {170329},
year = {2026},
issn = {0003-4916},
doi = {https://doi.org/10.1016/j.aop.2025.170329},
url = {https://www.sciencedirect.com/science/article/pii/S0003491625004117}
}

@article{SSLee,
  title = {Low-energy effective theory of {F}ermi surface coupled with {U}(1) gauge field in $2+1$ dimensions},
  author = {Lee, Sung-Sik},
  journal = {Phys. Rev. B},
  volume = {80},
  issue = {16},
  pages = {165102},
  numpages = {13},
  year = {2009},
  month = {Oct},
  publisher = {American Physical Society},
  doi = {10.1103/PhysRevB.80.165102},
  url = {http://link.aps.org/doi/10.1103/PhysRevB.80.165102}
}

@article{metlsach1,
  title = {Quantum phase transitions of metals in two spatial dimensions. {I}. {I}sing-nematic order},
  author = {Metlitski, Max A. and Sachdev, Subir},
  journal = {Phys. Rev. B},
  volume = {82},
  issue = {7},
  pages = {075127},
  numpages = {24},
  year = {2010},
  month = {Aug},
  publisher = {American Physical Society},
  doi = {10.1103/PhysRevB.82.075127},
  url = {http://link.aps.org/doi/10.1103/PhysRevB.82.075127}
}

@article{metlsach,
  title = {Quantum phase transitions of metals in two spatial dimensions. {II}. {S}pin density wave order},
  author = {Metlitski, Max A. and Sachdev, Subir},
  journal = {Phys. Rev. B},
  volume = {82},
  issue = {7},
  pages = {075128},
  numpages = {30},
  year = {2010},
  month = {Aug},
  publisher = {American Physical Society},
  doi = {10.1103/PhysRevB.82.075128},
  url = {http://link.aps.org/doi/10.1103/PhysRevB.82.075128}
}

@article{mross,
  title = {Controlled expansion for certain non-{F}ermi-liquid metals},
  author = {Mross, David F. and McGreevy, John and Liu, Hong and Senthil, T.},
  journal = {Phys. Rev. B},
  volume = {82},
  issue = {4},
  pages = {045121},
  numpages = {19},
  year = {2010},
  month = {Jul},
  publisher = {American Physical Society},
  doi = {10.1103/PhysRevB.82.045121},
  url = {http://link.aps.org/doi/10.1103/PhysRevB.82.045121}
}

@ARTICLE{Jiang,
   author = {{Jiang}, H.-C. and {Block}, M.~S. and {Mishmash}, R.~V. and 
	{Garrison}, J.~R. and {Sheng}, D.~N. and {Motrunich}, O.~I. and 
	{Fisher}, M.~P.~A.},
    title = "{Non-{F}ermi-liquid d-wave metal phase of strongly interacting electrons}",
  journal = {\nat},
     year = 2013,
    month = jan,
   volume = 493,
    pages = {39-44},
      doi = {10.1038/nature11732},
   adsurl = {http://adsabs.harvard.edu/abs/2013Natur.493...39J}
}

@article{Lee-Dalid,
  title = {Perturbative non-{F}ermi liquids from dimensional regularization},
  author = {Dalidovich, Denis and Lee, Sung-Sik},
  journal = {Phys. Rev. B},
  volume = {88},
  issue = {24},
  pages = {245106},
  numpages = {20},
  year = {2013},
  month = {Dec},
  publisher = {American Physical Society},
  doi = {10.1103/PhysRevB.88.245106},
  url = {http://link.aps.org/doi/10.1103/PhysRevB.88.245106}
}

@article{ips-fflo,
  title = {Non-{F}ermi liquid at the {FFLO} quantum critical point},
  author = {Pimenov, Dimitri and Mandal, Ipsita and Piazza, Francesco and Punk, Matthias},
  journal = {Phys. Rev. B},
  volume = {98},
  issue = {2},
  pages = {024510},
  numpages = {19},
  year = {2018},
  month = {Jul},
  publisher = {American Physical Society},
  doi = {10.1103/PhysRevB.98.024510},
  url = {https://link.aps.org/doi/10.1103/PhysRevB.98.024510}
}

@ARTICLE{ips-rahul,
       author = {{Mandal}, Ipsita and {Nandkishore}, Rahul M.},
        title = "{Interplay of Coulomb interactions and disorder in three-dimensional quadratic band crossings without time-reversal symmetry and with unequal masses for conduction and valence bands}",
      journal = {\prb},
         year = 2018,
        month = mar,
       volume = {97},
       number = {12},
          eid = {125121},
        pages = {125121},
          doi = {10.1103/PhysRevB.97.125121}
}

@article{moon-xu,
  title = {Non-Fermi-Liquid and Topological States with Strong Spin-Orbit Coupling},
  author = {Moon, Eun-Gook and Xu, Cenke and Kim, Yong Baek and Balents, Leon},
  journal = {Phys. Rev. Lett.},
  volume = {111},
  issue = {20},
  pages = {206401},
  numpages = {5},
  year = {2013},
  month = {Nov},
  publisher = {American Physical Society},
  doi = {10.1103/PhysRevLett.111.206401},
  url = {https://link.aps.org/doi/10.1103/PhysRevLett.111.206401}
}

@article{senshank,
  title = {{F}ermi Surfaces in General Codimension and a New Controlled Nontrivial Fixed Point},
  author = {Senthil, T. and Shankar, R.},
  journal = {Phys. Rev. Lett.},
  volume = {102},
  issue = {4},
  pages = {046406},
  numpages = {4},
  year = {2009},
  month = {Jan},
  publisher = {American Physical Society},
  doi = {10.1103/PhysRevLett.102.046406},
  url = {http://link.aps.org/doi/10.1103/PhysRevLett.102.046406}
}

@article{ips2,
  title = {Higher angular momentum pairing from transverse gauge interactions},
  author = {Chung, Suk Bum and Mandal, Ipsita and Raghu, Srinivas and Chakravarty, Sudip},
  journal = {Phys. Rev. B},
  volume = {88},
  issue = {4},
  pages = {045127},
  numpages = {7},
  year = {2013},
  month = {Jul},
  publisher = {American Physical Society},
  doi = {10.1103/PhysRevB.88.045127},
  url = {http://link.aps.org/doi/10.1103/PhysRevB.88.045127}
}

@article{ips3,
title = "{Pairing in half-filled Landau level}",
journal = "Annals of Physics ",
volume = "351",
number = "",
pages = "727 - 738",
year = "2014",
note = "",
issn = "0003-4916",
doi = "http://dx.doi.org/10.1016/j.aop.2014.09.021",
url = "http://www.sciencedirect.com/science/article/pii/S0003491614002814",
author = "Zhiqiang Wang and Ipsita Mandal and Suk Bum Chung and Sudip Chakravarty"
}

@Article{ips-sc,
  title = {Superconducting instability in non-{F}ermi liquids},
  author = {Mandal, Ipsita},
  journal = {Phys. Rev. B},
  volume = {94},
  issue = {11},
  pages = {115138},
  numpages = {15},
  year = {2016},
  month = {Sep},
  publisher = {American Physical Society},
  doi = {10.1103/PhysRevB.94.115138},
  url = {https://link.aps.org/doi/10.1103/PhysRevB.94.115138}
}

@article{ips-c2,
title = "Scaling behaviour and superconducting instability in anisotropic non-{F}ermi liquids",
journal = "Annals of Physics",
volume = "376",
pages = "89 - 107",
year = "2017",
issn = "0003-4916",
doi = "https://doi.org/10.1016/j.aop.2016.11.009",
url = "http://www.sciencedirect.com/science/article/pii/S0003491616302585",
author = "Ipsita Mandal"
}

@Article{ips-subir,
  title = {Hyperscaling violation at the {I}sing-nematic quantum critical point in two-dimensional metals},
  author = {Eberlein, Andreas and Mandal, Ipsita and Sachdev, Subir},
  journal = {Phys. Rev. B},
  volume = {94},
  issue = {4},
  pages = {045133},
  numpages = {16},
  year = {2016},
  month = {Jul},
  publisher = {American Physical Society},
  doi = {10.1103/PhysRevB.94.045133},
  url = {https://link.aps.org/doi/10.1103/PhysRevB.94.045133}
}

@ARTICLE{YBKim,
  title = {Theory of the nodal nematic quantum phase transition in superconductors},
  author = {Kim, Eun-Ah and Lawler, Michael J. and Oreto, Paul and Sachdev, Subir and Fradkin, Eduardo and Kivelson, Steven A.},
  journal = {Phys. Rev. B},
  volume = {77},
  issue = {18},
  pages = {184514},
  numpages = {9},
  year = {2008},
  month = {May},
  publisher = {American Physical Society},
  doi = {10.1103/PhysRevB.77.184514},
  url = {https://link.aps.org/doi/10.1103/PhysRevB.77.184514}
}

@ARTICLE{nayak,
       author = {{Nayak}, Chetan and {Wilczek}, Frank},
        title = "{Renormalization group approach to low temperature properties of a non-Fermi liquid metal}",
      journal = {Nuclear Physics B},
         year = 1994,
        month = nov,
       volume = {430},
       number = {3},
        pages = {534-562},
          doi = {10.1016/0550-3213(94)90158-9}
}

@article{ips-nfl-u1,
  title = {Critical {F}ermi surfaces in generic dimensions arising from transverse gauge field interactions},
  author = {Mandal, Ipsita},
  journal = {Phys. Rev. Research},
  volume = {2},
  issue = {4},
  pages = {043277},
  numpages = {17},
  year = {2020},
  month = {Nov},
  publisher = {American Physical Society},
  doi = {10.1103/PhysRevResearch.2.043277},
  url = {https://link.aps.org/doi/10.1103/PhysRevResearch.2.043277}
}

@article{hsu2008superconductivity,
author = {Fong-Chi Hsu  and Jiu-Yong Luo  and Kuo-Wei Yeh  and Ta-Kun Chen  and Tzu-Wen Huang  and Phillip M. Wu  and Yong-Chi Lee  and Yi-Lin Huang  and Yan-Yi Chu  and Der-Chung Yan  and Maw-Kuen Wu },
title = {Superconductivity in the {P}b{O}-type structure $\alpha$-{F}e{S}e},
journal = {Proceedings of the National Academy of Sciences},
volume = {105},
number = {38},
pages = {14262-14264},
year = {2008},
doi = {10.1073/pnas.0807325105},
URL = {https://www.pnas.org/doi/abs/10.1073/pnas.0807325105}
}

@article{margadonna2008crystal,
  author ="Margadonna, Serena and Takabayashi, Yasuhiro and McDonald, Martin T. and Kasperkiewicz, Karolina and Mizuguchi, Yoshikazu and Takano, Yoshihiko and Fitch, Andrew N. and Suard, Emmanuelle and Prassides, Kosmas",
title  ={Crystal structure of the new {F}e{S}e$_{1-x}$ superconductor},
journal  ="Chem. Commun.",
year  ="2008",
issue  ="43",
pages  ="5607-5609",
publisher  ="The Royal Society of Chemistry",
doi  = {https://doi.org/10.1039/b813076k},
url  = {http://dx.doi.org/10.1039/B813076K}
}

@article{mcqueen2009tetragonal,
  title = {Tetragonal-to-Orthorhombic Structural Phase Transition at 90 {K} in the Superconductor {F}e$_{1.01}${S}e},
  author = {McQueen, T. M. and Williams, A. J. and Stephens, P. W. and Tao, J. and Zhu, Y. and Ksenofontov, V. and Casper, F. and Felser, C. and Cava, R. J.},
  journal = {Phys. Rev. Lett.},
  volume = {103},
  issue = {5},
  pages = {057002},
  numpages = {4},
  year = {2009},
  month = {Jul},
  publisher = {American Physical Society},
  doi = {10.1103/PhysRevLett.103.057002},
  url = {https://link.aps.org/doi/10.1103/PhysRevLett.103.057002}
}

@article{ips-zero-mode,
  title = {Zero sound and plasmon modes for non-{F}ermi liquids},
journal = {Physics Letters A},
volume = {447},
pages = {128292},
year = {2022},
issn = {0375-9601},
doi = {https://doi.org/10.1016/j.physleta.2022.128292},
url = {https://www.sciencedirect.com/science/article/pii/S0375960122003747},
author = {Ipsita Mandal}
}

@article{ips-kazi,
title = {Generic deformation channels for critical Fermi surfaces in the collisionless regime},
journal = {Annals of Physics},
volume = {457},
pages = {169409},
year = {2023},
issn = {0003-4916},
doi = {https://doi.org/10.1016/j.aop.2023.169409},
url = {https://www.sciencedirect.com/science/article/pii/S0003491623001951},
author = {Kazi Ranjibul Islam and Ipsita Mandal}
}

@book{kadanoff2018quantum,
  title={Quantum statistical mechanics},
  author={Kadanoff, Leo P},
  year={2018},
  publisher={CRC Press}
}

@article{craig,
    author = {Craig, R. A.},
    title = {Perturbation Expansion for Real‐Time {G}reen's Functions},
    journal = {Journal of Mathematical Physics},
    volume = {9},
    number = {4},
    pages = {605-611},
    year = {1968},
    month = {04},
    issn = {0022-2488},
    doi = {10.1063/1.1664616},
    url = {https://doi.org/10.1063/1.1664616}
    }

@article{rammer1986quantum,
  title = {Quantum field-theoretical methods in transport theory of metals},
  author = {Rammer, J. and Smith, H.},
  journal = {Rev. Mod. Phys.},
  volume = {58},
  issue = {2},
  pages = {323--359},
  numpages = {0},
  year = {1986},
  month = {Apr},
  publisher = {American Physical Society},
  doi = {10.1103/RevModPhys.58.323},
  url = {https://link.aps.org/doi/10.1103/RevModPhys.58.323}
}
\end{document}